\documentclass[preprint,showpacs,amsmath,amssymb]{revtex4}

\usepackage{graphicx}
\usepackage{dcolumn}
\usepackage{bm}
\newcommand{\be}{\begin{equation}}
\newcommand{\ee}{\end{equation}}
\newcommand{\bea}{\begin{eqnarray}}
\newcommand{\eea}{\end{eqnarray}}
\newcommand{\ba}{\begin{array}}
\newcommand{\ea}{\end{array}}
\newcommand{\bi}{\begin{itemize}}
\newcommand{\ei}{\end{itemize}}

\newcommand{\refe}[1]{(\ref{#1})}

\begin{document}

\title{Coupled-channel analysis of $K\Sigma$ production on the nucleon up to 2.0 GeV}

\author{Xu Cao{\footnote{Electronic address: Xu.Cao@theo.physik.uni-giessen.de}} }
\author{V. Shklyar{\footnote{Electronic address: Vitaliy.Shklyar@theo.physik.uni-giessen.de}}}
\author{H. Lenske{\footnote{Electronic address: Horst.Lenske@theo.physik.uni-giessen.de}}}

\affiliation{Institut f\"{u}r Theoretische Physik, Universit\"{a}t Giessen, D-35392 Giessen, Germany}

\begin{abstract}

A coupled-channel effective Lagrangian model respecting unitary and gauge invariance is applied to
the combined analysis of the $(\pi,\gamma) N \to K\Sigma$ reactions for the center of mass energies up to 2 GeV. The recent photoproduction data obtained by the CLAS, CBELSA, LEPS, and GRAAL groups are included into our calculations with the aim to extract the resonance couplings to the $K\Sigma$ state. Both resonances and background contributions are found to be important to reproduce a correct shape of the angular distributions and polarization observables. Our description to the data is of good quality. The extracted properties of isospin $I = 3/2$ resonances are discussed in detail while the $I = 1/2$ resonances are largely determined by the non-strangeness channels.

\end{abstract}
\maketitle{}

\section{INTRODUCTION}

Strangeness production on the nucleon has attracted a lot of attention since a long time ago. It is not only an elementary process of strangeness production, but also an ideal place to look for the resonances that might be weakly coupled to the $\pi N$ state. Recently, the interest in the $K \Sigma$ channel has been revived by the new photoproduction data with improved precision gained by several experimental groups including LEPS~\cite{LEPS03,LEPS06Sumihama,LEPS06Kohri}, CLAS~\cite{CLASthesis,CLAS04McNabb,CLAS06sigma0,CLAS07CxCz,CLAS10Dey,CLAS10sigmap}, CBELSA~\cite{CBELSA08,CBELSA12}, GRAAL07~\cite{GRAAL07}, and SAPHIR~\cite{SAPH04,SAPH05}. Strangeness electroproduction on the proton is also accurately measured by the CLAS~\cite{CLASelectro} and A1 Group~\cite{A1electro}. However, on theoretical side, most of the calculations~\cite{Mart1995,Mart2000,Janssen02,Saghai1995,ZPLi1995} are based on the previous database and the very recent data are not yet included into the analysis except for the Bonn-Gatchina partial wave analysis~\cite{BoCh05,BoCh05theo,BoCh07,BoCh08,BoCh11,BoCh12}. So it is meaningful to perform a full coupled-channel calculation based on the updated database combining both the $\pi N$ and $\gamma N$ data. The Giessen K-matrix model is ready for this kind of analysis\cite{Feuster98,Feuster99,Penner:2002a,Penner:2002b,Penner:thesis}. Results on the non-strangeness production in an updated version of the model have been published in a series of papers~\cite{Shklyar04J,Shklyar05ome,Shklyar07eta,Shklyar12eta}. The partial wave amplitudes of $K \Lambda$ production and the coupling strength of resonances to the K$\Lambda$ channel have also been extracted in Ref.~\cite{Shklyar05lam}. Herein, we give a coherent coupled-channel analysis of the $K \Sigma$ production in the Giessen model.

Another motivation of the present paper is to resolve the current inconclusive status of different models on strangeness production, especially in the $K \Sigma$ channel. In the isobar model of Refs~\cite{Mart1995,Mart2000}, the $K^0 \Sigma^+$ photoproduction besides the $K^+ \Sigma^0$ channel was found to be important for extracting the knowledge on the background contributions. The $P_{13}(1720)$ resonance was shown to be essential to describe the data of this channel. In another isobar model~\cite{Janssen02} it was pointed out that the bare Born terms largely overestimated the data. In this calculation, five resonances, i.e. $S_{11}(1650)$, $P_{11}(1710)$, $P_{13}(1720)$, $S_{31}(1900)$, and $P_{31}(1910)$ were found to be sufficient for achieving a good agreement with data. Only little or no evidence for a $D_{13}(1895)$ state was found although that state seemed to emerge in the previous studies of the $K\Lambda$ photoproduction~\cite{MartLambda}. In Ref.~\cite{Saghai1995}, another very comprehensive isobar model was built for a combined analysis of photo- and electro-production data. It included nucleonic resonances with spin up to $J = 5/2$, hyperonic resonances with spin $J = 1/2$, and kaonic resonances. A chiral quark model~\cite{ZPLi1995} with the spin $J \leq 7/2$ resonances in the s-channel found that the contact term and the resonances with isospin $I = 3/2$, i.e. the $F_{37}(1950)$, $F_{35}(1905)$, $P_{33}(1920)$, and $P_{31}(1910)$, were dominant in the $K \Sigma$ photoproduction. After these very early isobar models which were used to analyze the old data before 2002 (for data references, see~\cite{Saghai1995,ZPLi1995}), a Regge-plus-resonance (RPR) model~\cite{Corthals07} was promoted to describe the new LEPS, CLAS, GRAAL and SAPHIR data. Its background terms were deduced from the high-energy Regge-trajectory exchange in the t-channel. Only four isospin $I = 1/2$ resonances, namely the $S_{11}(1650)$, $P_{11}(1710)$, $P_{13}(1720)$ and $P_{13}(1900)$, and four isospin $I = 3/2$ resonances, namely the $S_{31}(1900)$, $P_{31}(1910)$, $D_{33}(1700)$ and $P_{33}(1920)$ were needed to describe the data.

However, the coupled-channel analysis of the $K \Sigma$ photoproduction is, in fact, rather scarce. The early Giessen model analysis~\cite{Penner:2002a,Penner:2002b,Penner:thesis} included resonances with spin up to $J = 3/2$ and obtained a fair agreement with the old $K\Lambda$ and $K\Sigma$ photoproduction data. A similar coupled-channel model with the K-matrix approach firstly developed by Usov\&Scholten~\cite{Usov2005} and later extended by Shyam et al.~\cite{Shyam10} considered also resonances with spins up to $J = 3/2$ and fitted its parameters to the SAPHIR data~\cite{SAPH04,SAPH05}. Different gauge-restoration procedures were compared and the Davidson-Workman prescription, also being used in the Giessen model~\cite{Penner:2002b}, was found to work best.

The Juelich group made a coupled-channel analysis of $\pi^+ p \to K^+ \Sigma^+$~\cite{Doring11} which is a pure isospin $I = 3/2$ channel. The selection of final states was recently expanded to other $K \Sigma$ charged states, together with the $\eta N$ and $K \Lambda$ channels~\cite{Doring12}. It was extended to the $\pi N$ photoproduction~\cite{Huang12}, but has not been employed to analyze the strangeness photoproduction. A dynamical coupled-channel formalism developed by Juli\'{a}-D\'{i}az et al.~\cite{Julia2006} used a chiral constituent quark model for strangeness photoproduction and investigated recent data on the $K \Lambda$ photoproduction combining with the $\pi^- p \to K \Lambda$ and $K^0\Sigma^0$ data. Including limited isospin $I = 3/2$ resonances, they found three new resonances: a $D_{13}$, a $S_{11}$ and a $P_{13}$ with mass around 1954MeV, 1806 MeV and 1893 MeV, respectively. On the other hand it is not clear whether these resonances play a role in the $K \Sigma$ photoproduction. A chiral unitary framework addressing the importance of gauge-invariance was developed in Ref.~\cite{Borasoy07} but only focused on the close-to-threshold region of strangeness photoproduction due to the difficulty in dealing with the higher chiral orders.

Very recently the CLAS and CBELSA groups have released a lot of accurate data~\cite{CBELSA08,CBELSA12,CLAS10Dey,CLAS10sigmap} so enlarged considerably the database of strangeness production on the nucleon. Especially, the $\gamma p \to K^0 \Sigma^+$ data published by the CBELSA group~\cite{CBELSA08,CBELSA12} are much more precise than the old SAPHIR data~\cite{SAPH05}. An interesting and important conclusion to be drawn from the CBELSA data is that most of the previous calculations overestimated the total cross section of this channel. While the two $\gamma p \to K^+ \Sigma^0$ datasets published respectively by the CLAS~\cite{CLAS06sigma0} and SAPHIR Collaboration~\cite{SAPH04} are not very consistent in the backward angles, the newly measured data by the CLAS group~\cite{CLAS10Dey} agree well with the former CLAS data~\cite{CLAS06sigma0} and LEPS data~\cite{LEPS03,LEPS06Sumihama,LEPS06Kohri}. In the Bonn-Gatchina isobar partial wave analysis of these data, the evidence for the $P_{13}(1900)$ resonance which is not favored by diquark models is reported~\cite{BoCh07,BoCh08,BoCh11,BoCh12} while the $P_{31}(1750)$ state found both in the Juelich~\cite{Doring11} and Giessen coupled-channel model~\cite{Penner:2002a,Penner:2002b} plays no role. Keeping these problems in mind, we perform a new combined analysis by taking into account all new measurements from CLAS and CBELSA groups etc.

We start in Sec.~\ref{formalism} with a brief outline of the main features of the Giessen model.
The detailed calculations of the $\pi N\to K \Sigma$ and $\gamma N\to K \Sigma$ and the extracted resonance parameters are presented in Sec.~\ref{fitparam}. We finish with a short summary in Sec.~\ref{summary}.

\section{Giessen Model} \label{formalism}

\begin{figure}
  \begin{center}
{\includegraphics*[width=14cm]{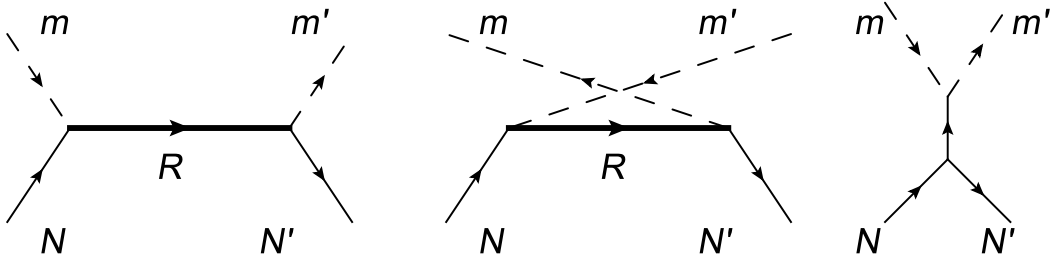}}
       \caption{
$s$-, $u$- and $t$- channel contributions to the interaction potential. The $m$ and $m'$ stand for
initial and final mesons. The $N$, $N'$ and $R$ label initial, final and intermediate baryons. Time flows from left to the right.
      \label{diagr}}
  \end{center}
\end{figure}

Though QCD is established as a theory of the strong interaction for a long time, only effective degrees of freedom --- mesons and baryons --- are observed in experiment. Based on this observation, we develop a coupled-channel unitary Lagrangian model to study the reaction mechanism of the pion- and photo-induced reactions in the resonance region. The details of the interaction Lagrangians in the model and results for the non-strange channels can be found in Ref.~\cite{Penner:2002a,Penner:2002b,Penner:thesis} and Ref.~\cite{Shklyar04J,Shklyar05ome,Shklyar07eta,Shklyar12eta} respectively. Here, we only briefly outline the main ingredients of our model for simplicity.  In order to obtain the scattering amplitude $T_{fi}$, the Bether-Salpeter equation is solved in the $K$-matrix approximation where the real part of the propagator $G_{ab}$ is neglected:
\bea
T_{fi}=K_{fi}+i\sum_{a,b}K_{fa}\emph{Im}(G_{ab})T_{bi},
\label{Kapp}
\eea
with $i, f$ and $a(b)$ being the initial, final and intermediate states, respectively. The equation $K_{fi} = V_{fi} + \sum_{a,b} V_{fa}Re(G_{ab})K_{bi}$ is reduced to be $K = V$ in terms of the $K$-matrix approximation. In this way the solution of the multichannel problem becomes feasible, satisfying the important condition of unitarity\cite{Penner:2002a,Penner:2002b,Penner:thesis}. The validity of this approximation has been discussed in Ref.~\cite{Penner:2002a,BoCh12,Shklyar04J}. A proper theoretical definition of the resonance parameters would be to perform the calculation in the complex energy plane and search directly for the eigenstates of the underlying Hamiltonian amounting to determine the eigenvalues as poles on the second Riemann-sheet. However, these kind of calculations are numerically quite involved and beyond the scope of our present work. Here we continue the previous efforts ~\cite{Feuster98,Feuster99,Penner:2002a,Penner:2002b,Penner:thesis,Shklyar04J,Shklyar05ome,Shklyar07eta,Shklyar12eta,Shklyar05lam} to use K-matrix approximation and quantify resonances by their input Breit-Wigner (BW) masses and widths. When compared to the values of Particle Data Group (PDG), the BW mass and widths are also quoted from the PDG's publication~\cite{pdg2010}.

After completing the partial-wave decomposition in terms of Wigner $d$-functions, the Bether-Salpeter equation finally reduces to a set of  algebraic equations for the scattering $T$-matrix~\cite{Penner:thesis}:
\bea
T^{J\pm,I}_{fi} = \left[\frac{ K^{J\pm,I}}{1-iK^{J\pm,I}}\right]_{fi},
\label{GBS3}
\eea
where $J^{\pm}$,$I$ are total spin, parity and isospin of the final and initial states \mbox{$f,i =$ $ \gamma N$}, $\pi N$, $2\pi N$, $\eta N$, $\omega N$, $K\Lambda$, $K\Sigma$. The experimental observables, i.e. the cross sections and polarization observables, could be directly calculated by the $T^{J\pm,I}_{fi}$, as explicitly expressed in the Appendix G of Ref.\cite{Penner:2002a} for the pion-induced reactions and in the Appendix E of Ref.\cite{Penner:2002b} for the photo-induced reactions. A graphically presentation of physical meaning and measurement of polarization observables could be found in Ref.~\cite{CLAS06sigma0}. In our model the $2\pi$ state is described in terms of the effective isovector-scalar meson. This allows to control the $2\pi N$ inelastic flux and fix the resonance couplings to the $2\pi N$ channel~\cite{Penner:thesis}.

The interaction potential ($K$-matrix) is constructed as a sum of the  $s$-, $u$-
and $t$-channel  tree-level Feynman diagrams as depicted in Fig.~\ref{diagr}. It is calculated from the corresponding effective interaction Lagrangians which respect chiral symmetry in low-energy regime\cite{Penner:2002a,Penner:2002b}. To cut off the contributions from large four-momenta $q^2\gg \Lambda^2$, each meson and baryon vertex is dressed by a corresponding form factor of the form:
\bea
F_p (q^2,m^2) &=& \frac{\Lambda^4}{\Lambda^4 +(q^2-m^2)^2}.
\label{formfact}
\eea
We use the same cutoffs for all resonances with given spin $J$, e.g., $\Lambda^{N^*(1535)}_{i}$=$\Lambda^{N^*(1650)}_{j}$ where indices $i$,$j$ run over all final states. We also choose the cutoff at the $NK\Sigma$ vertex with the same value of the nucleon cutoff: $\Lambda_{NK\Sigma}$=$\Lambda_{N}$ = 0.95 GeV. Hence, the number of free parameters is largely decreased.

The non-resonant part of the transition amplitude of $(\pi,\gamma)N \to K\Sigma$
is the same as used in the previous Giessen model studies\cite{Penner:2002a,Penner:2002b,Shklyar05lam}.
It consists of the nucleon Born term and $t$-channel contributions
with the $K^*$, $K^*_0$, and $K_1$ mesons in the intermediate state. It should be mentioned that $t$-channel $K_1$ meson exchange only contributes to the $\gamma N \to K \Sigma$ and $K^*_0$ meson exchange only to $\pi N \to K \Sigma$. The $K^*$ meson exchange contributes to both reactions. The couplings $g_{K^* K \pi}$, $g_{K^*_0 K \pi}$, $g_{{K^*} K \gamma}$ and $g_{K_1 K \gamma}$ are calculated from the experimental decay widths of the PDG compilation~\cite{pdg2010}, see~\cite{Penner:2002b} for the values. We use the same $\Lambda_t$ = 0.77 GeV at the corresponding $t$-channel vertices for both associated strangeness and non-strangeness channels \cite{Shklyar05lam}. Similar to our previous studies~\cite{Penner:2002a,Penner:2002b} we do not consider the $u$-channel diagrams to the $(\pi,\gamma) N \to K\Sigma$ reaction in oder to keep the model as simple as possible. The calculation of such kind of diagrams would require a priori unknown couplings to the intermediate strange baryons, i.e. the $\Lambda^*$ and $\Sigma^*$ resonances.

We treat the photoproduction reactions perturbatively owing to the smallness of the electromagnetic coupling . This means that the summation on intermediate states runs only over hadronic states by neglecting $\gamma N$ state in Eq.~\refe{GBS3}. This prescription has been checked in \cite{Penner:2002b} and found to be very accurate.

\section{Numerical RESULTS AND DISCUSSION} \label{fitparam}

In our calculation we include 11 isospin $I = 1/2$ resonances and 9 isospin $I = 3/2$ resonances, listed in Tab.~\ref{resonance12} and Tab.~\ref{resonance32}, respectively. The effects of the isospin $I = 1/2$ resonances have been extensively studied in the production of $\omega N$~\cite{Shklyar05ome}, $\eta N$\cite{Shklyar07eta,Shklyar12eta} and $K\Lambda$\cite{Shklyar05lam} by including spin $J \leq 5/2$ resonances~\cite{Shklyar04J}. The high spin resonances are found to be important in the $\omega N$ production~\cite{Shklyar05ome}. A discussion on the $I = 1/2$ partial waves in the elastic $\pi N \to \pi N$, the proton and neutron multipoles of $\gamma N \to \pi N$, the $\pi N \to 2\pi N$ total partial wave cross sections and the $\pi N$ inelasticity can be found in Ref.~\cite{Shklyar05ome}. In this paper we continue the investigations of the $I = 1/2$ and $3/2$ sectors with the parameters fitted to newly published $K\Sigma$ photoproduction data together with the previous $\pi N \to K\Sigma$ measurements (for data references, see e.g ~\cite{Penner:2002a,Penner:2002b}) in the energy region $\sqrt{s} \leq 2.0$ GeV. The included $K\Sigma$ photoproduction data are those of the $\gamma p \to K^+ \Sigma^0$ published by the LEPS~\cite{LEPS03,LEPS06Sumihama,LEPS06Kohri}, CLAS~\cite{CLAS06sigma0,CLAS10Dey} and GRAAL~\cite{GRAAL07} group, and those of $\gamma p \to K^0 \Sigma^+$ released by the CLAS~\cite{CLASthesis} and CBELSA~\cite{CBELSA08} collaboration, respectively. The SAPHIR data have been left out here because of the already mentioned inconsistencies of the $K^+ \Sigma^0$ data~\cite{SAPH04} with the corresponding CLAS and GRAAL data (for the details, see Ref.~\cite{CLAS10Dey}). Also, the $K^0 \Sigma^+$ SAPHIR data~\cite{SAPH05} have much bigger error bars than those of the CBELSA and CLAS group. Here, the data before 2002 are also no longer used. The Giessen model results for these old data base can be found in our previous publications~\cite{Penner:2002a,Penner:2002b}.

In the present calculation we achieve a quite satisfactory description of the $\gamma p \to K^+ \Sigma^0$ data ($\chi^2 = 1.8$) and the $\gamma p \to K^0 \Sigma^+$ data ($\chi^2 = 2.0$). However, the pion-induced strangeness production reactions are described slightly less accurate as indicated by the corresponding $\chi^2$ values of $\chi^2 =$ 4.1, 3.2 and 2.8 for the $\pi^+ p \to K^+ \Sigma^+$, $\pi^- p \to K^0 \Sigma^0$ and $\pi^- p \to K^+ \Sigma^-$ reactions, respectively. The parameters that have been varied in our fit simultaneously to the $I = 1/2$ and $3/2$ sectors are shown in Tabs.~\ref{resonance12} -~\ref{borncoupling}. Due to the smallness of the $N^* K \Sigma$ couplings, all previously obtained BW masses, branching ratios and couplings corresponding to non-strangeness production~\cite{Shklyar05ome} are hardly affected by the additional $K\Sigma$ photoproduction data, so in the following subsection we will concentrate on the properties of the $I = 3/2$ resonances.

\subsection{Partial wave analysis in the isospin $3/2$ sector} \label{pwa32new}

\begin{figure}
  \begin{center}
{\includegraphics*[width=17cm]{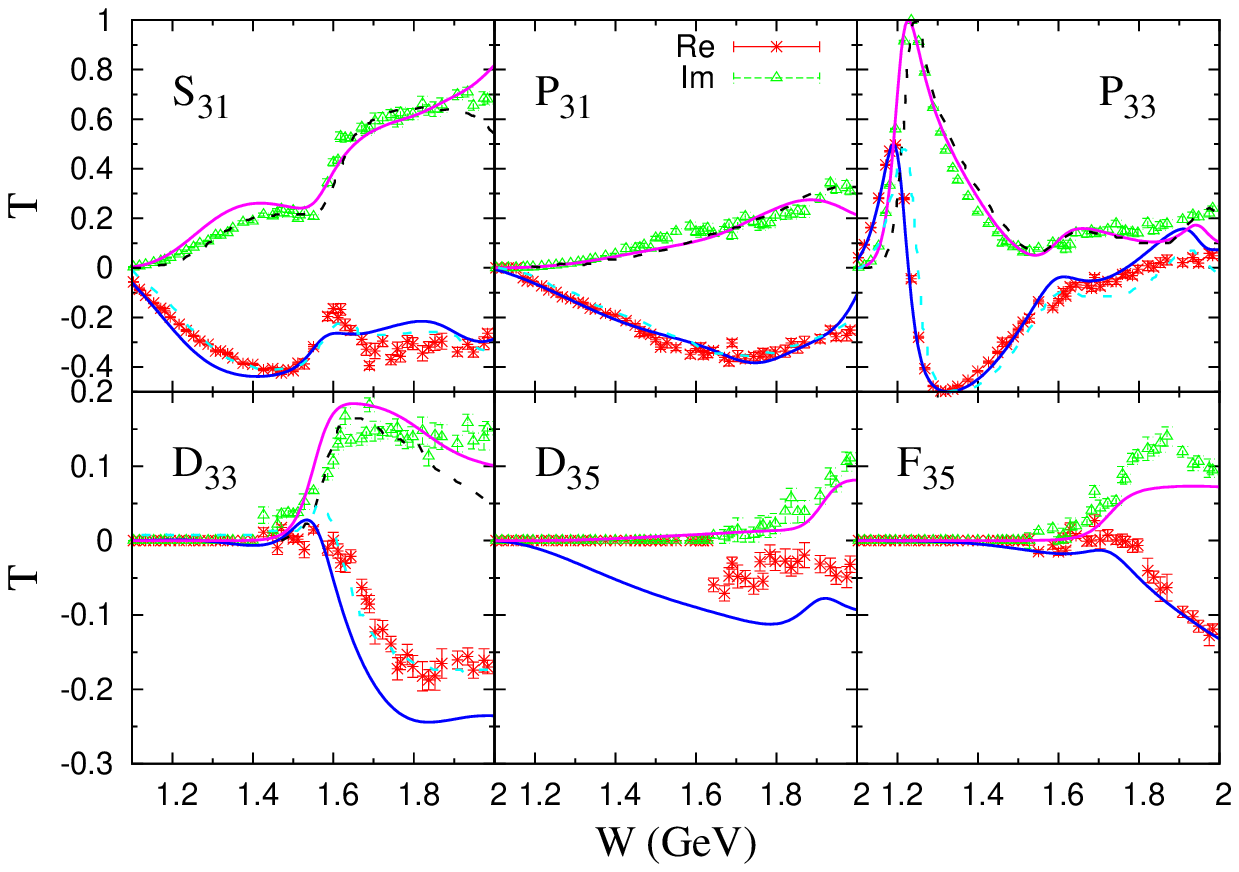}}
       \caption{
(Color online) The elastic $\pi N$ partial waves for $I = 3/2$. The upper solid (magenta) lines, dashed (dark grey) lines and triangle (green) points are imaginary part of amplitude of our model, previous Giessen model~\cite{Penner:2002a}, and GWU/SAID analysis~\cite{Arndt2006}, respectively. The lower solid (blue) lines, dashed (cyan) lines and star (red) points are the correspondent real part of amplitude.
      \label{piNpwa}}
  \end{center}
\end{figure}

\begin{figure}
  \begin{center}
{\includegraphics*[width=17cm]{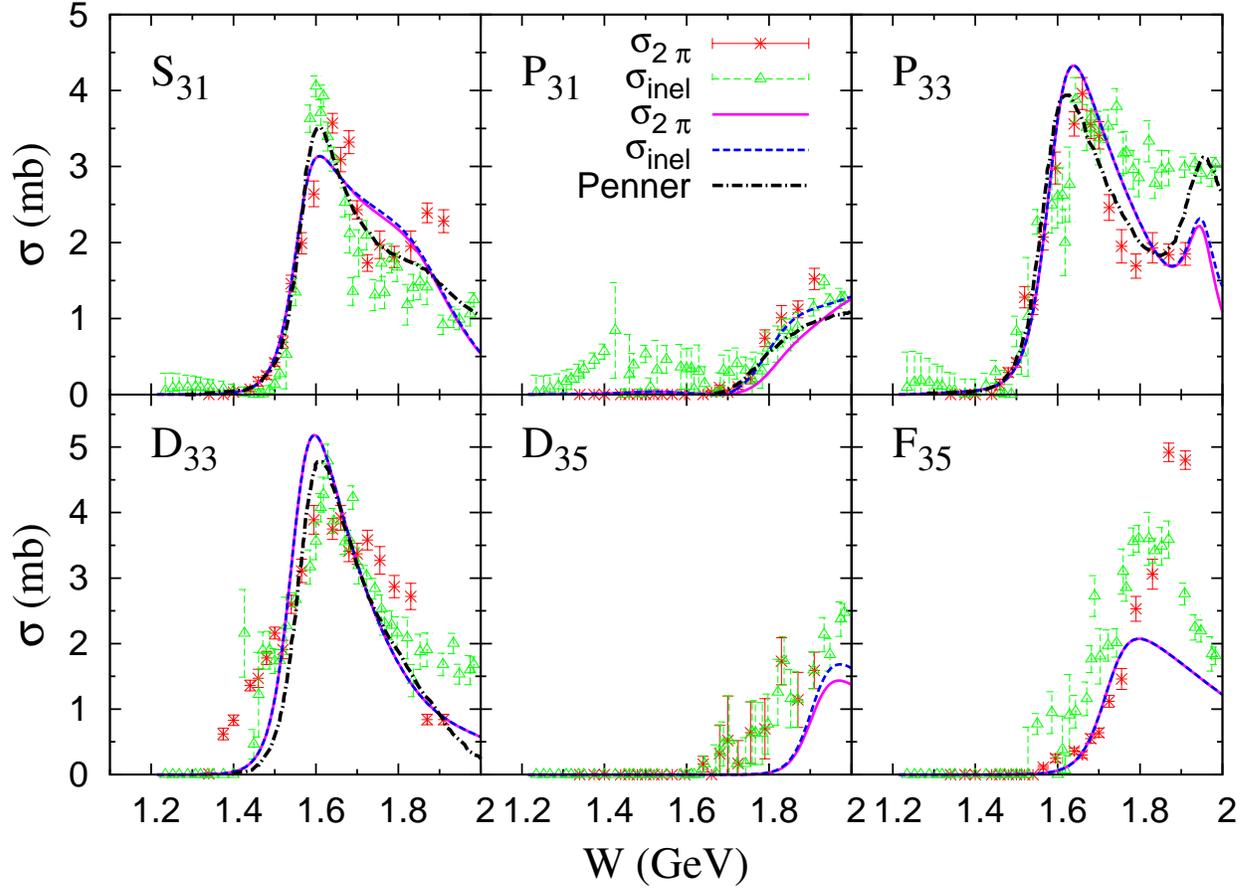}}
       \caption{
(Color online) The inelastic $2\pi N$ total partial wave cross sections for $I = 3/2$. The solid lines (magenta) and dashed line (blue) are the calculated cross section and inelasticity, respectively. The dash-dotted (dark grey) lines are the calculated cross section of previous Giessen model~\cite{Penner:2002a}. The triangle points (green) and star points (red) are the total cross section and inelasticity from Manley et al.~\cite{Arndt1984} and GWU group~\cite{Arndt2006}, respectively. In $S_{31}$, $D_{33}$ and $F_{35}$ waves, the calculated inelasticities almost coincide with the calculated $2\pi N$ cross sections.
      \label{pi2pwa}}
  \end{center}
\end{figure}

\begin{figure}
  \begin{center}
{\includegraphics*[width=17cm]{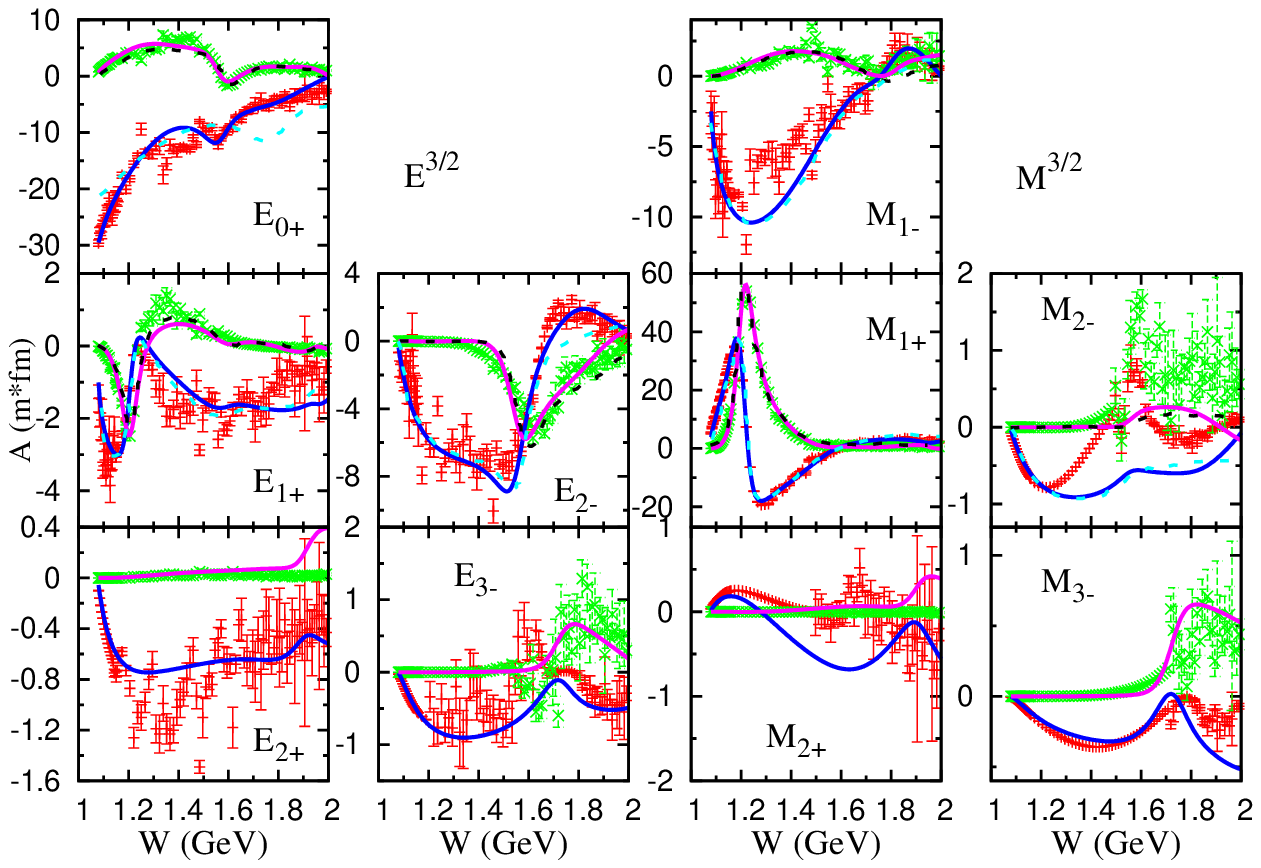}}
       \caption{
(Color online) The $\gamma N \to \pi N$ multipoles for $I = 3/2$.  The upper solid (magenta) lines, dashed (dark grey) lines and triangle (green) points are imaginary part of amplitude of our model, previous Giessen model~\cite{Penner:2002a}, and GWU analysis~\cite{Arndt2002}, respectively. The lower solid (blue) lines, dashed (cyan) lines and star (red) points are the correspondent real part of amplitude.
      \label{gammaNpwa}}
  \end{center}
\end{figure}

The parameters of the isospin $I = 3/2$ resonances used in our calculations were extensively discussed in the previous publications~\cite{Penner:2002a,Penner:2002b,Shklyar04J}. Here we will comment on the new features after adding resonances with spin $J = 5/2$ and updating our database. The calculated isospin $I = 3/2$ partial waves of $\pi N \to \pi N$, $\pi N \to 2\pi N$ and multipoles of $\gamma N \to \pi N$ are shown in Fig.~\ref{piNpwa}, Fig.~\ref{pi2pwa} and Fig.~\ref{gammaNpwa}, respectively. In the following we compare our parameters of $I = 3/2$ resonances with those of other coupled-channel models, i.e. the previous Giessen model~\cite{Penner:2002a,Penner:2002b}, the Usov\&Scholten~\cite{Usov2005,Shyam10}, and the Juelich model~\cite{Doring11}, and those of partial wave analyses, i.e. the Bonn-Gatchina model~\cite{BoCh05,BoCh11,BoCh05theo,BoCh07,BoCh08,BoCh12}, the KSU~\cite{Manley1992,Manley2012}, the Pitt-ANL~\cite{Vrana2000} and the GWU~\cite{Arndt1995,Arndt2006,Arndt1984,Arndt2002,Arndt2012}.

\textbf{$S_{31}$ partial wave:} The first $S_{31}$ resonance with mass around 1620 MeV is well identified in many analyses due to the obvious structures of $S_{31}$ partial waves of the $\pi N \to \pi N$ and $\pi N \to 2 \pi N$, and also the unambiguous $E^{3/2}_{0+}$ multipole in this energy region~\cite{Arndt2002,Arndt2012}. Our present fit of this multipole is much better than that in the old Giessen calculations where an obvious deviation between theory and data appears between 1.4 and 1.6 GeV~\cite{Penner:2002a}, see Fig.~\ref{gammaNpwa}. However, the description of this elastic $\pi N$ partial wave become worse as shown in Fig.~\ref{piNpwa}, which indicates that the model faces a demanding challenge from the new $K\Sigma$ photoproduction data. The $S_{31}$-wave inelasticity mainly comes from $2\pi N$ channel as can be seen from Fig.~\ref{pi2pwa}.

The second $S_{31}$ resonance around 1900 MeV is controversial and only of 2-star status in PDG ranking~\cite{pdg2010}. It is found in the partial wave analyses of KSU~\cite{Manley1992,Manley2012} and Pitt-ANL~\cite{Vrana2000} but not confirmed by the GWU survey~\cite{Arndt1995,Arndt2006}. The Bonn-Gatchina group previously concluded that it did not need to include this resonance~\cite{BoCh05,BoCh07,BoCh08}. However they do add it to their recent analysis~\cite{BoCh11,BoCh12} and find that its properties are consistent with those obtained in the recent KSU analysis~\cite{Manley2012}. In the previous Giessen calculations, the $S_{31}(1900)$ emerges only in the Pascalutsa prescription of the $J = 3/2$ resonances vertices~\cite{Penner:2002a,Penner:2002b}, but the evidence is weak and rather of non-resonant nature. In our present calculations we re-investigated the relevance of this resonance and find a small electromagnetic helicity amplitude $A_{1/2}$ with a value around $-10 \times 10^{-3}$ GeV$^{-1/2}$. In the region close to 2 GeV, this $S_{31}(1900)$ creates small structures in the $S_{31}$ waves of $\pi N$ elastic and $2\pi N$ channel and in the $E^{3/2}_{0+}$ multipole of $\pi N$ photoproduction which are still within the experimental uncertainties. But for $K \Sigma$ production it is irrelevant, being totally emersed into the background. The contribution to the cross section of that state can be fully compensated by a small variation of the Born couplings. So we do not include this second $S_{31}$ resonance here. Other coupled-channel models, i.e. Juelich and Usov\&Scholten include only $S_{31}(1620)$, too. Our extracted BW mass and width of $S_{31}(1620)$ are respectively 1598 MeV and 209 MeV which are close to the values of the recent KSU analysis~\cite{Manley2012}.

\textbf{$P_{31}$ partial wave:} In our model only one $P_{31}(1750)$ is sufficient to reproduce a correct shape in the partial waves of $\pi N$ and $2\pi N$ channels, though its electromagnetic properties have large uncertainty because of the big error bars in the $M^{3/2}_{1-}$ multipoles~\cite{Arndt2002,Arndt2012} as can be seen in Fig.~\ref{gammaNpwa}. The Fig.~\ref{pi2pwa} shows that the $P_{31}$ inelasticity comes mainly from $2 \pi N$ channel but also from the $K\Sigma$ final state with the branching ratio Br$(P_{31}(1750) \to K\Sigma )\simeq0.90\%$ as shown in Tab.~\ref{resonance32}. The Pitt-ANL~\cite{Vrana2000} and old KSU models~\cite{Manley1992} include the additional $P_{31}(1910)$ resonance and find the $P_{31}(1750)$ with the mass at about 1730 MeV, which is a little lower than ours. The GWU analysis~\cite{Arndt1995,Arndt2006} finds a $P_{31}$ pole around 1770 MeV but attributes it to the $P_{31}(1910)$ resonance due to its large mass which is above 2000 MeV and much higher than the value in other models. In the Juelich approach~\cite{Doring11} the $P_{31}(1750)$ state is dynamically generated and the $P_{31}(1910)$ state is a genuine resonance. The recent KSU~\cite{Manley2012} and Bonn-Gatchina~\cite{BoCh12} model also found only the $P_{31}(1910)$ resonance but no evidence for the $P_{31}(1750)$ state. The Usov\&Scholten calculation~\cite{Usov2005,Shyam10} does not include any $P_{31}$ resonance.

\textbf{$P_{33}$ partial wave:} In our analysis, we need three resonances in this partial wave: $P_{33}(1232)$, $P_{33}(1600)$, and $P_{33}(1920)$. This selection of states is widely confirmed by different groups, i.e. Pitt-ANL, KSU and Bonn-Gatchina. Especially the properties of  $P_{33}(1232)$ are well established because of the accurate $E^{3/2}_{1+}$ and $M^{3/2}_{1+}$ multipole data at this energy, as shown in Fig.~\ref{gammaNpwa}. The $P_{33}(1920)$ whose BW mass is found to be above 2 GeV is needed to deliver additional $\pi N$ strength in this partial wave at high energies. But in the GWU model only the first two $P_{33}$ resonances are included into the previous fit~\cite{Arndt1995}, and their recent calculation take only the first resonance into account~\cite{Arndt2006}. In the Juelich model~\cite{Doring11} the $P_{33}(1600)$ state is a dynamically generated state, while the other two $P_{33}$ states are of genuine character. The $P_{33}(1600)$ is not present in the analysis of Usov\&Scholten~\cite{Usov2005,Shyam10} but the other two states, the $P_{33}(1232)$ and $P_{33}(1920)$, are included into their model. Here it is worth to mention that the $P_{33}(1600)$ is found to be important in the double-pion production in nucleon-nucleon collisions, especially in the $pp \to nn\pi^+\pi^+$ channel where only isospin $I = 3/2$ resonances can contribute~\cite{Caotwopi}. The authors of Ref.~\cite{Caotwopi} obtain 350 MeV for the BW width of $P_{33}(1600)$.

In fact, as seen in Fig.~\ref{piNpwa} there is no clear resonance structure in the elastic $\pi N$ partial wave at high energies, so it is a little difficult to establish the existence of $P_{33}(1600)$ and $P_{33}(1920)$ unambiguously. For these two higher lying $P_{33}$ states, we find small electromagnetic contributions as shown in Tab.~\ref{resonance32}, so they have hardly any visible structure in the $M^{3/2}_{1+}$ and $E^{3/2}_{1+}$ multipoles. Also, the large error bars in the $E^{3/2}_{1+}$ multipole at high energies inhibit solid conclusion.

As already addressed in previous Giessen model analyses~\cite{Penner:2002a}, the $2 \pi N$ channel is dominant in the $P_{33}$-wave inelasticity but the observed inelastic partial-wave cross section is higher than our model calculation by about 1 mb above 1.7 GeV, as can be clearly seen in Fig.~\ref{pi2pwa}. As a result, we still miss inelastic contributions of about 1 mb in the $P_{33}$ partial wave at high energies in the present model, similarly as the case in the $P_{13}$ wave~\cite{Penner:2002a}. This is a possible hint for the contribution of a $3 \pi N$ state such as $\rho \Delta$.

\textbf{$D_{33}$ partial wave:} In this partial wave we find only the $D_{33}(1700)$ resonance and, in this respect, agree with other models except the old KSU analysis~\cite{Manley1992} where the second $D_{33}(1940)$ was found. However, the recent KSU investigation~\cite{Manley1992} does not find this $D_{33}(1940)$ resonance so it seems that now all analyses are converging to only one $D_{33}$ state. The $D_{33}$ wave in the $\pi N$ elastic channel shows a distinct resonance structure as can be seen in Fig.~\ref{piNpwa} so the BW mass of $D_{33}(1700)$ is fixed with good accuracy. The agreement between the calculated real part of the $D_{33}$ partial wave and the GWU single energy elastic $\pi N$ solution~\cite{Arndt2002,Arndt2012} becomes a little worse after the inclusion of the new $K \Sigma$ photoproduction data, as shown in Fig.~\ref{piNpwa}.

As in the previous Giessen model investigation~\cite{Penner:2002a}, the calculated $2 \pi N$ inelasticity does not follow well the results of Manley et al.~\cite{Arndt1984} below the $D_{33}(1700)$ resonance position, as shown in Fig.~\ref{pi2pwa}. As discussed in the case of $P_{33}$ wave, this would be amended by an extension of the model into the $3 \pi N$ sector.

In our previous studies~\cite{Penner:2002b}, there were also discrepancies in the description of the $M^{3/2}_{2-}$ multipole and it is difficult to extract accurate helicity amplitudes there. It was supposed that this was due to the lack of background contribution in this multipole. The present results demonstrate a better agreement for the imaginary part but the strength of the real part is still not big enough to explain the data, as depicted in Fig.~\ref{gammaNpwa}. The trend of the $E^{3/2}_{2-}$ multipole is nicely reproduced, though small deviations are seen at the intermediate energies.

\textbf{$D_{35}$ partial wave:} In this channel we include the $D_{35}(1930)$ resonance. Our present calculations have a problem in the description of this amplitude, as seen in Fig.~\ref{piNpwa}. In the previous fit to only pion-induced reactions we have found that it is difficult to get a reasonable agreement with the GWU results for this partial wave if only $D_{35}(1930)$ state is taken into account~\cite{Shklyar04J}. Being extended into the photo-induced reactions, our model demonstrates a better result for the imaginary part of the $D_{35}$ wave than the situation for the purely hadronic results, as can be seen from Fig.~\ref{piNpwa}. It is interesting to note that the similar problems are experienced in the Juelich model~\cite{Doring11}. In addition, the calculated $2 \pi N$ cross section tends to be below the results of Manley et al.~\cite{Arndt1984} as shown in Fig.~\ref{pi2pwa}, which might point out to some deficiency in the description of the  $2\pi N$ channel.

In Fig.~\ref{gammaNpwa} some deviations from the GWU analysis can be seen in the real part of the $M^{3/2}_{2+}$ amplitude but there are large error bars in this multipoles of the GWU analysis~\cite{Arndt2002,Arndt2012}. The background constitutes the main contribution to the $M^{3/2}_{2+}$ and $E^{3/2}_{2+}$ multipoles and the $D_{35}(1930)$ resonance has small structures at high energies due to its small electromagnetic couplings.

Other models also include this $D_{35}(1930)$ with the only exception of the Bonn-Gatchina model which does not consider any $D_{35}$ resonance. A second $D_{35}(2350)$ is included in the Pitt-ANL and the old GWU calculations, but recent GWU analyses find no indication for this resonance. In our model we do not consider this high mass state because of its minor contribution to the energy region below 2 GeV.

\textbf{$F_{35}$ partial wave:} In our previous hadronic result it was sufficient to take a single $F_{35}(1905)$ resonance into account for a good description of this partial wave~\cite{Shklyar04J}. In the present analysis, the imaginary part of amplitude of the elastic $\pi N$ channel in Fig.\ref{piNpwa} and the partial wave cross section of $2\pi N$ channel in Fig.\ref{pi2pwa} obviously are both underestimated above 1.8 GeV, so other inelastic channels such as $3 \pi N$ may also contribute. The $F_{35}(1905)$ state contributes also to the $E^{3/2}_{3-}$ and $M^{3/2}_{3-}$ multipoles, as can be seen in Fig.~\ref{gammaNpwa}. But here we include another $F_{35}(2000)$ resonance, though its BW mass is close to the upper energy limit of our calculation. The reason is that it considerably improves the high energy tail of our $S_{31}$, $P_{31}$ and $P_{33}$ partial waves in the elastic $\pi N$ channels. However, the signal of this state is hard to resolve unambiguously because of its small partial decay width to $\pi N$ channel.

The existence of $F_{35}(1905)$ is confirmed by many other studies because of its obvious role in the $F_{35}$ wave of the $\pi N$ elastic scattering. Another $F_{35}$ resonance with a lower mass of about 1750 MeV found in the old KSU study~\cite{Manley1992} is not needed in our calculations. The ambiguity of $F_{35}(2000)$ is still unresolved. Though the latest GWU~\cite{Arndt2006}, Bonn-Gatchina~\cite{BoCh11,BoCh12} analyses as well as many other former analyses find no evidence for this resonance, the recent KSU survey~\cite{Manley2012} finds it with mass and width of 2015(24) MeV and 500(52) MeV, respectively, which has to be compared to our values of 2160 MeV and 313 MeV, listed in Tab.~\ref{resonance32}.

\subsection{Results for the $\pi N \to K \Sigma$ reaction}  \label{KSigmapi}

Our calculated total cross sections are compared to the available data in Fig.~\ref{pstots}. As can be seen, when both isospin $I = 1/2$ and $I = 3/2$ channels are accessible, the $S_{11}$ wave dominates at threshold. Other partial waves, namely $P_{31}$, $P_{33}$, $D_{35}$ and $D_{15}$, contribute at high energies. The effect from $D_{13}$ and $D_{33}$ waves is hardly seen. When only isospin $I = 3/2$ is allowed, i.e the $\pi^+ p \to K^+ \Sigma^+$ channel, the $P_{31}$ wave dominates at threshold and the $D_{35}$ and $P_{33}$ waves become important at high energies. At the very close-to-threshold region, the contribution of the $S_{31}$ wave is noticeable. The $F_{15}$ and $F_{35}$ partial waves tend to be negligible in whole energy range.

The non-resonant part of the amplitude in the $\pi^- p \to K^+ \Sigma^-$ reaction is negligibly small but it is seen in the $\pi^- p \to K^0 \Sigma^0$ and $\pi^+ p \to K^+ \Sigma^+$ channels. This is because the contribution from the nucleon Born term is very small and the non-resonant contribution comes mainly from the t-channel $K_0^*$ meson exchange. In our results the coupling constant of $NK^*\Sigma$ is much smaller than that in our previous investigations, see Tab~\ref{borncoupling}. As a result, the contribution from the t-channel $K^*$ meson exchange is reduced. Though effects of the non-resonant part of the amplitude affect the $\Sigma$-polarization, its overall contribution is very small.

It should be stressed that the exact shape of angular distributions and polarization observables is produced by the interference of several partial waves. So sometimes even contributions of small magnitude will influence the shape significantly. Therefore it is necessary to look deeper into the reaction amplitudes, including also the weakly populated reaction channels. In the following two subsections we would concentrate on the differential cross sections and polarization observables of the $\pi^+ p \to K^+ \Sigma^+$ and $\pi^- p \to K^+ \Sigma^-/K^0 \Sigma^0$ reactions, respectively.

\subsubsection{Results for the $\pi^+ p \to K^+ \Sigma^+$ reaction} \label{KSigmaplus}

\begin{figure}
  \begin{center}
{\includegraphics*[width=17cm]{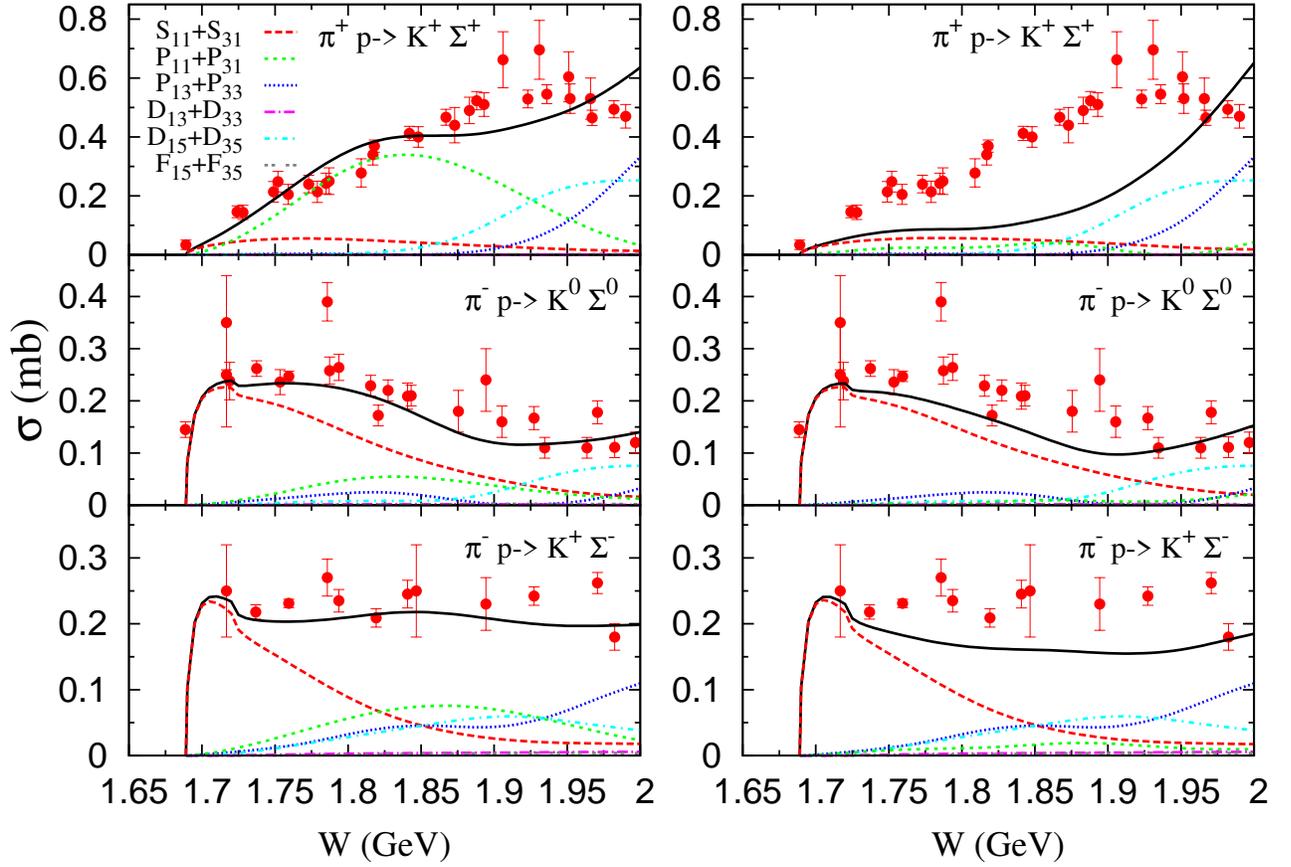}}
       \caption{
(Color online) The total cross sections of $\pi N \to K \Sigma$. Left panel: the full model calculation. Right panel: the model calculation with $P_{31}(1750)$ turned off. The solid (black), dashed (red), short dashed (green),  dotted (blue),  dot dashed (magenta), dot short dashed (cyan), and double short dashed (grey) correspond to the full results, the $S_{I1}$ ($\frac{1}{2}^-$), $P_{I1}$ ($\frac{1}{2}^+$), $P_{I3}$ ($\frac{3}{2}^+$), $D_{I3}$ ($\frac{3}{2}^-$), $D_{I5}$ ($\frac{5}{2}^-$), and $F_{I5}$ ($\frac{5}{2}^+$), respectively.
      \label{pstots}}
  \end{center}
\end{figure}

\begin{figure}
  \begin{center}
{\includegraphics*[width=17cm]{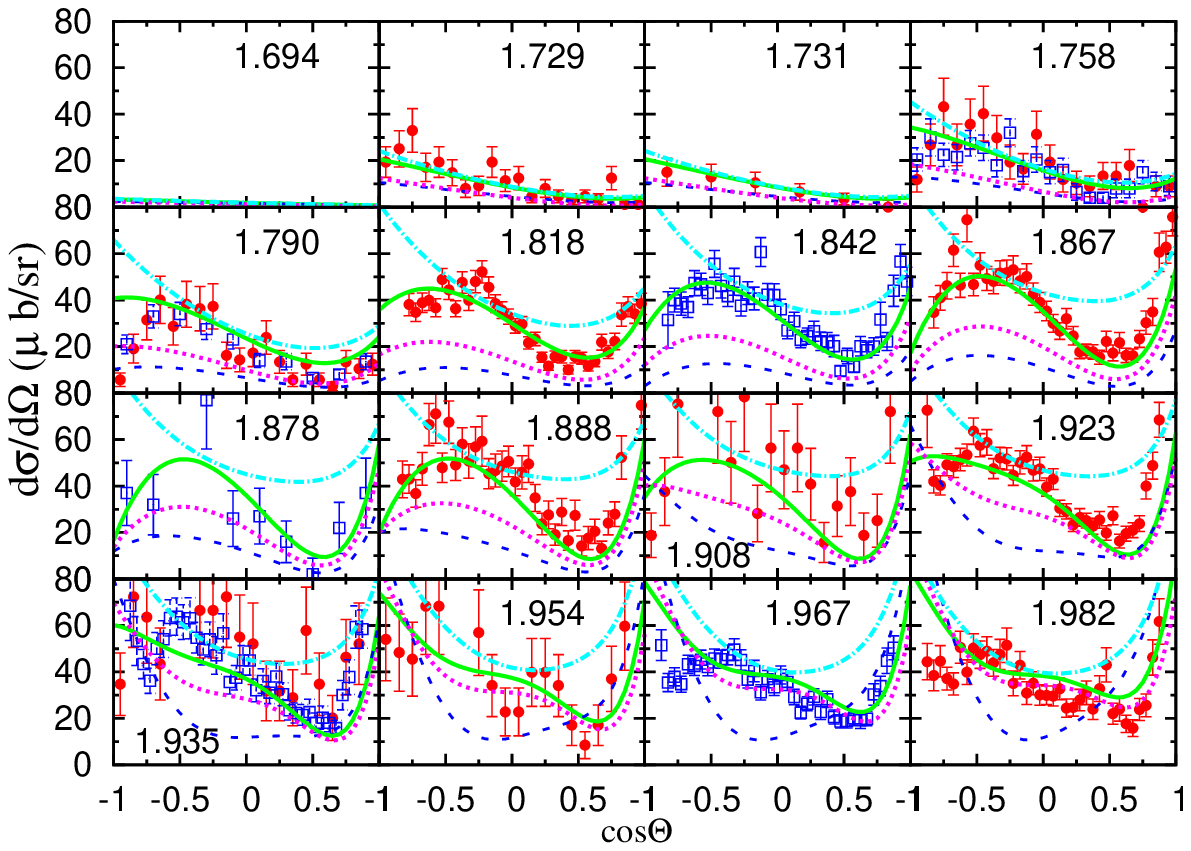}}
       \caption{
(Color online) The differential cross section of $\pi^+ p \to K^+ \Sigma^+$ reaction. The solid (green), dashed (blue), dotted (magenta) and dash-dotted (cyan) lines are the full model calculation, the model calculation with the $P_{31}(1750)$, $D_{33}(1700)$ and $D_{35}(1930)$ turned off, respectively. The numeric values label the center of mass energies in unit of GeV.
      \label{pspdif}}
  \end{center}
\end{figure}

\begin{figure}
  \begin{center}
{\includegraphics*[width=17cm]{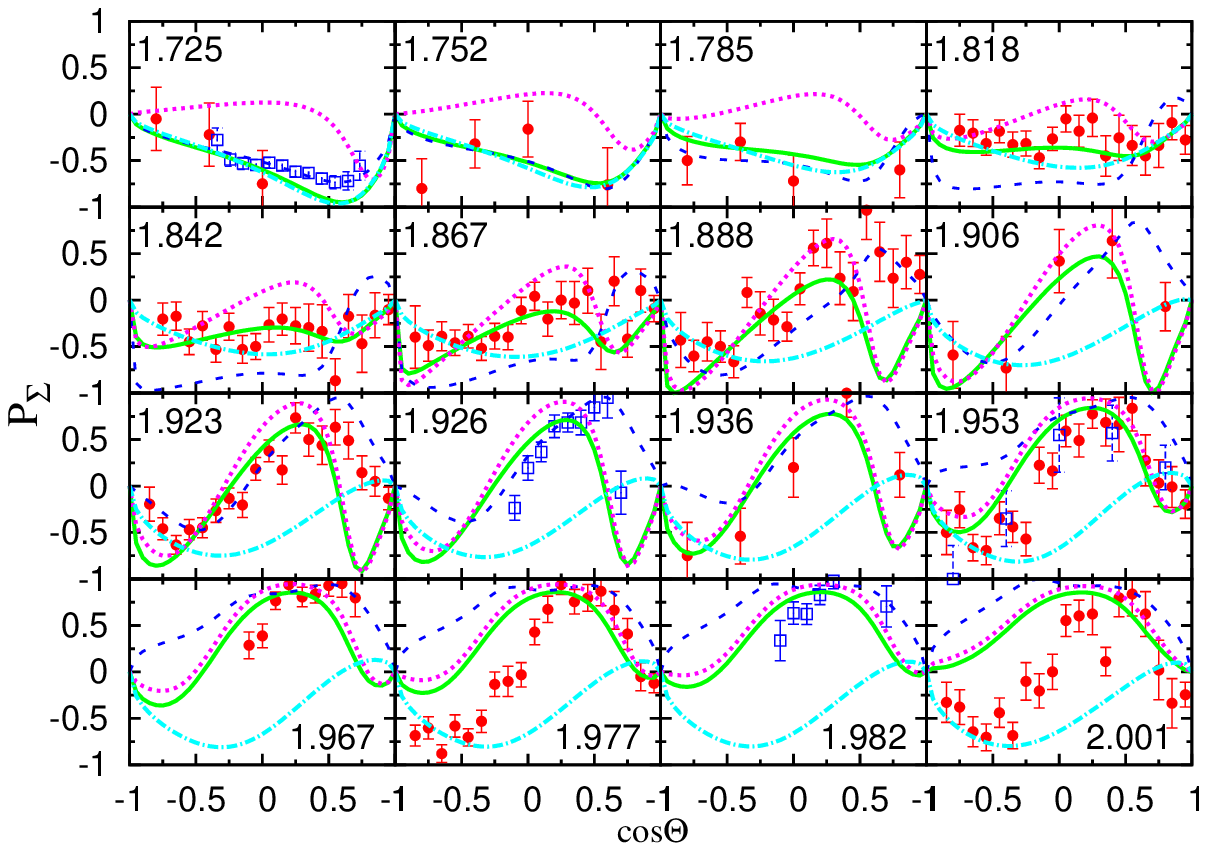}}
       \caption{
(Color online) The $\Sigma$-polarization of $\pi^+ p \to K^+ \Sigma^+$ reaction. The meaning of line is the same with that in Fig.~\ref{pspdif}. The numeric values label the center of mass energies in unit of GeV.
      \label{psprec}}
  \end{center}
\end{figure}

The $\pi^+ p \to K^+ \Sigma^+$ channel is purely isospin $I = 3/2$. Our conclusion on this channel is similar to the previous Giessen model study. In the region very close to threshold, the $S_{31}$ wave is dominated by the $S_{31}(1620)$ resonance whereas the $S_{11}$-wave contribution is forbidden. In the considered energy region, our present calculations show that the shape of the $\pi^+ p \to K^+ \Sigma^+$ angular distributions is dominated by the $P_{31}(1750)$ resonance together with the $P_{33}(1600)$, $D_{33}(1700)$ and $D_{35}(1930)$ states, and especially enhancing the strength of a broad peak at the backward angles, as shown in Fig.~\ref{pspdif}. For the $\Sigma$-polarization in Fig.~\ref{psprec}, the $P_{31}(1750)$ resonance is important at all energies and the $D_{33}(1700)$ is essential already close to threshold. It should be mentioned that the $D_{35}(1930)$ is responsible for the dip at forward angles above 1.8 GeV in the angular distribution (see Fig.~\ref{pspdif}) and the steep rise at intermediate angles seen in the $\Sigma$-polarization (see Fig.~\ref{psprec}). It seems that this result resolves the apparent confusion in the $D_{35}$ wave of elastic $\pi N$ collisions mentioned in Sec.~\ref{pwa32new}: the $D_{35}(1930)$ resonance is definitely needed for a good description of the $\pi^+ p \to K^+ \Sigma^+$ data. Similar to the other two charged channels, the transition current flows from $P_{31}(1750)$ into the $D_{35}$ and $P_{33}$ partial wave at high energies. As a result, these two partial waves exceed other partial wave contributions to be the strongest above 1.95 GeV, inducing a steep rise of the total cross section in Fig.~\ref{pstots}. In Fig.~\ref{pspdif} and Fig.~\ref{psprec}, we also show results of calculations where the $P_{31}(1750)$ resonance was turned off. The effect from $F_{35}(1905)$ and $F_{35}(2000)$ can be seen in the angular distributions and $\Sigma$-polarization though the overall contribution from these states is found to be small.

However, the conclusions vary much in different models. In the Juelich model~\cite{Doring11}, the $\pi^+ p \to K^+ \Sigma^+$ reaction is dominated by the $S_{31}(1620)$ while the $P_{33}$-wave dominated by the $P_{33}(1600)$ and $P_{33}(1920)$ resonances is ranking the second but much weaker than the contribution of the $S_{31}(1620)$ state. At the energies around 2.0 GeV, the $F_{37}(1950)$ begins to exceed the $P_{33}$ contribution but still is smaller than $S_{31}$ wave. In the Bonn-Gatchina analysis of this reaction~\cite{BoCh11}, the $P_{33}(1920)$ is identified to be the most essential and the interference of $J = 7/2^+$ and $J = 5/2^+$ channels plays an important role at high energies. So it seems that the $K^+ \Sigma^+$ production mechanism needs further clarification. Juli\'{a}-D\'{i}az et al.~\cite{Julia2006} achieve a reasonable agreement with the $\pi^- p \to K^0 \Sigma^0$ data by including only three isospin $I = 3/2$ resonances $S_{31}(1900)$, $P_{31}(1910)$, and $S_{33}(1920)$ together with several isospin $I = 1/2$ resonances.

\subsubsection{Results for the $\pi^- p \to K^+ \Sigma^-$ and $K^0 \Sigma^0$ reactions} \label{KSigmaneu}

\begin{figure}
  \begin{center}
{\includegraphics*[width=17cm]{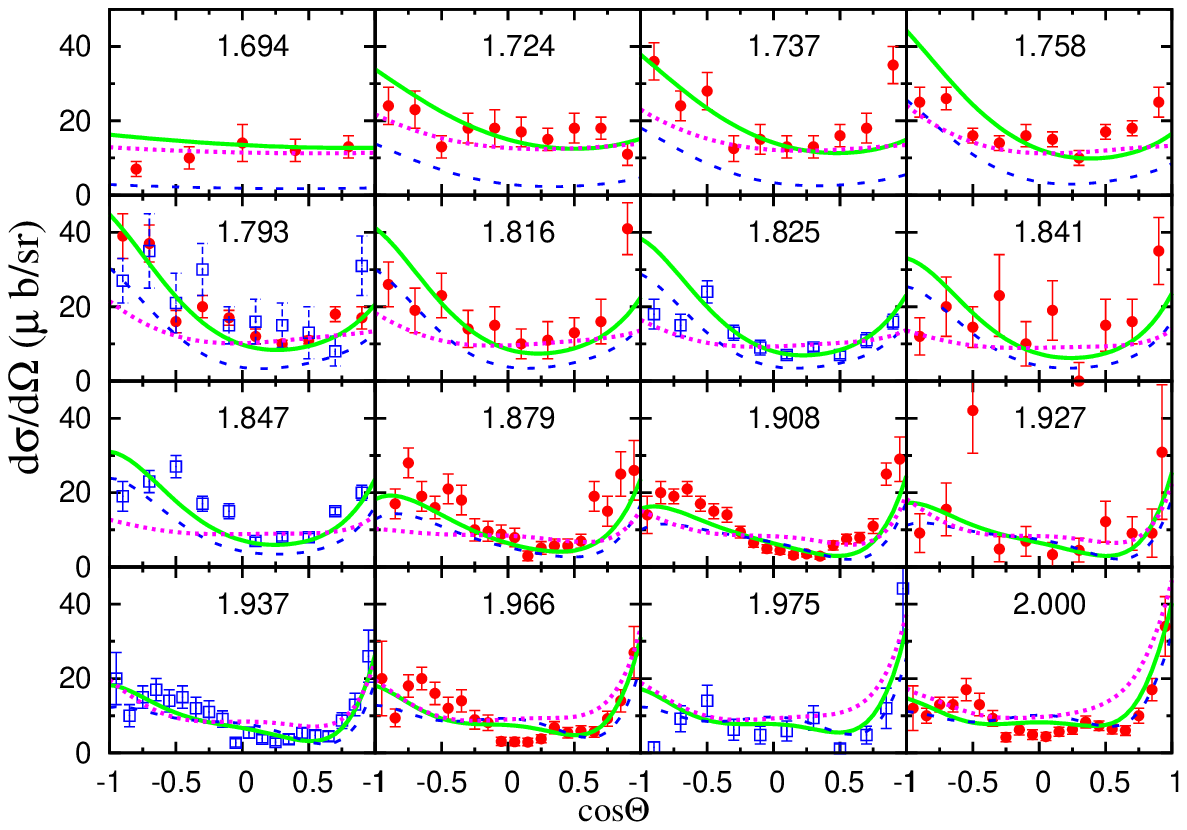}}
       \caption{
(Color online) The differential cross section of $\pi^- p \to K^0 \Sigma^0$ reaction. The solid (green), dashed (blue) and dotted (magenta) lines are the full model calculation, the model calculation with the $S_{11}(1650)$ and $F_{15}(1680)$ turned off, respectively. The numeric values label the center of mass energies in unit of GeV.
      \label{ps0dif}}
  \end{center}
\end{figure}

\begin{figure}
  \begin{center}
{\includegraphics*[width=17cm]{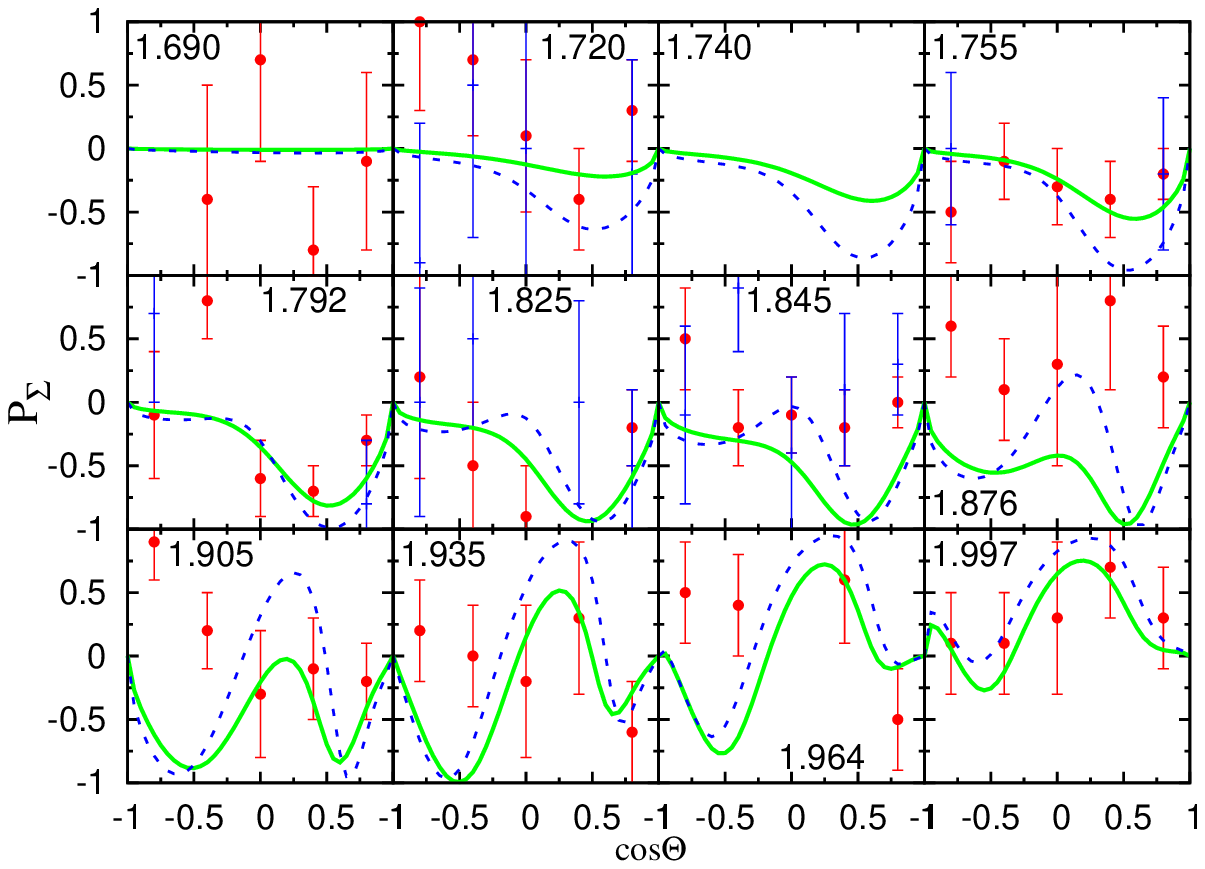}}
       \caption{
(Color online) The $\Sigma$-polarization of $\pi^- p \to K^0 \Sigma^0$ reaction. The meaning of line is the same with that in Fig.~\ref{ps0dif}. The numeric values label the center of mass energies in unit of GeV.
      \label{ps0rec}}
  \end{center}
\end{figure}

\begin{figure}
  \begin{center}
{\includegraphics*[width=17cm]{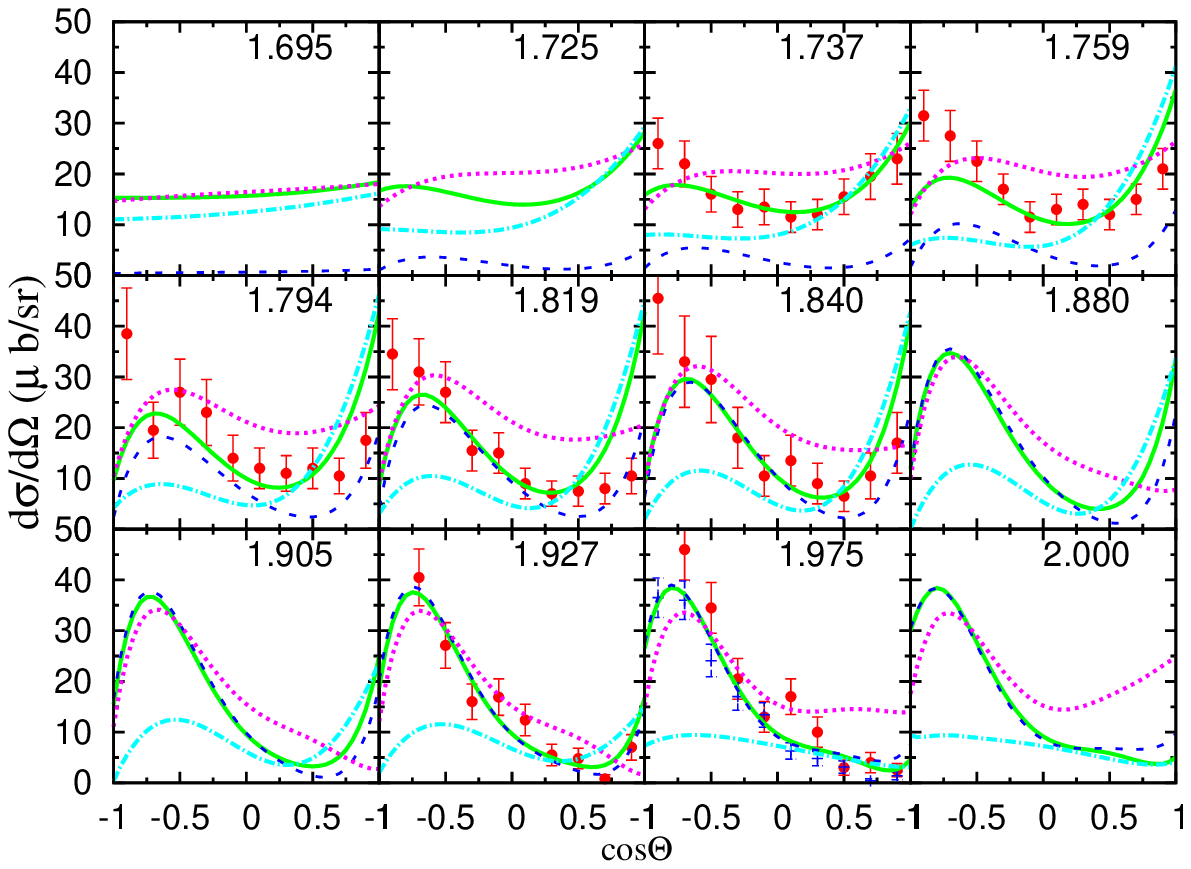}}
       \caption{
(Color online) The differential cross section of $\pi^- p \to K^+ \Sigma^-$ reaction. The solid (green), dashed (blue), dotted (magenta) and dash-dotted (cyan) lines are the full model calculation, the model calculation with the $S_{11}(1650)$, $D_{15}(1675)$ and $F_{15}(1680)$ turned off, respectively. The numeric values label the center of mass energies in unit of GeV.
      \label{psmdif}}
  \end{center}
\end{figure}

In the $\pi^- p \to K^+ \Sigma^-$ and $K^0 \Sigma^0$ channels, from threshold up to 1.8 GeV, the $S_{11}(1650)$ resonance dominates the energies and its destructive interference with the $S_{11}(1535)$ is also important. Close to threshold the effect from $S_{31}(1620)$ resonance is seen but much smaller than that of the $S_{11}(1650)$ state. The small kink at around 1.72 GeV is caused by the $\omega N$ threshold effect. At higher energies, the contributions of $P_{31}(1750)$, $P_{11}(1440)$, $P_{33}(1600)$ and $D_{35}(1930)$ become comparable. The important role of $D_{33}(1700)$ and $D_{15}(1675)$ states is clearly visible in angular distributions of various observables over the whole considered energy region, though they are small in the total cross sections. Such a behavior indicates that the enhancement is due to the interference effects which obviously are removed in the total cross sections by the angular integration. The $D_{35}$ and $P_{33}$ waves which mainly originate from the $D_{35}(1930)$ become significant in the energy tail while the contribution of $P_{31}(1750)$ decreases steadily. Our results are different from our previous investigation~\cite{Penner:2002a}, where the $S_{11}(1650)$ resonance dominates in the close-to-threshold region but the $P_{11}(1710)$ and $P_{31}(1750)$ are the strongest contribution for increasing energies. As we have pointed out, the impact from the $J = 5/2$ resonances, which were not included into the previous Giessen model~\cite{Penner:2002a}, is visible. In Fig.~\ref{ps0dif}, Fig.~\ref{ps0rec} and Fig.~\ref{psmdif}, we compare our full results to the calculations where the $S_{11}(1650)$ resonance was turned off in order to illustrate the role of this resonance.

In the $\pi^- p \to K^+ \Sigma^-$ and $\pi^- p \to K^0 \Sigma^0$ reactions, the contribution of the $P_{31}(1750)$ resonance is suppressed and it is only seen in the $\Sigma$-polarization of the $\pi^- p \to K^0 \Sigma^0$ channel. Instead, $S_{11}(1650)$, $P_{33}(1600)$, $D_{15}(1675)$, $D_{35}(1930)$ and $F_{15}(1680)$ all together determine the shape of the angular distributions and $\Sigma$-polarization of these two reactions. The $D_{15}(1675)$ state is responsible for the steep forward rise in angular distributions of the $\pi^- p \to K^+ \Sigma^-$ reaction at the intermediate energies, as seen in Fig.~\ref{psmdif}. The $F_{15}(1680)$ is significant in the backward structures in the angular distributions of both channels, as respectively shown in Fig.~\ref{ps0dif} and Fig.~\ref{psmdif}. The contribution of the $F_{15}(2000)$ resonance is very small but still noticeable.

\subsection{Results for the $\gamma p \to K \Sigma$ reaction}  \label{KSigmaphoto}

\begin{figure}
  \begin{center}
{\includegraphics*[width=17cm]{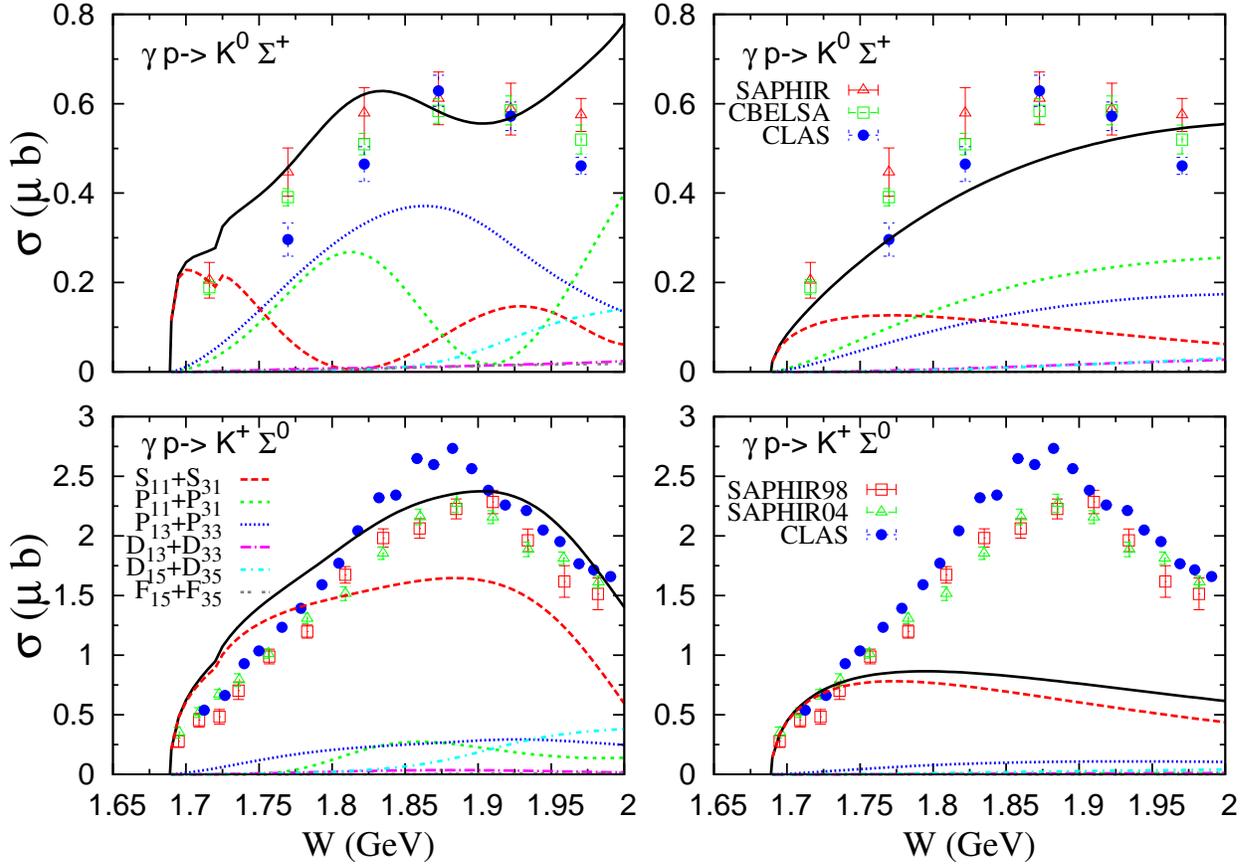}}
       \caption{
(Color online) The total cross sections of $\gamma N \to K \Sigma$. Left panel: the full model calculation. Right panel: the Born terms and t-channel meson exchange only. The line type is the same as in Fig.~\ref{pstots}. The data are taken from SAPHIR~\cite{SAPH04,SAPH05}, CBELSA~\cite{CBELSA08} and CLAS~\cite{CLAS06sigma0,CLASthesis}. The SAPHIR data is only for comparison but not used in the parameters fitting.
      \label{gstots}}
  \end{center}
\end{figure}

As depicted in the lower panels of Fig.~\ref{gstots}, the total cross section of the $\gamma p \to K^+ \Sigma^0$ reaction is largely dominated by the $S_{I1}$ channel while the contribution from other partial waves is small. In the case of the $\gamma p \to K^0 \Sigma^+$ channel the situation is much more complicated. The $S_{I1}$, $P_{I1}$ and $P_{I3}$ partial wave amplitudes are the most important ones. The $S_{I1}$ component has a second maximum around 1.93 GeV which is induced by the interference between the resonances and background generated by t-/u-channel interactions. However, in the finally obtained scattering amplitudes the two types of dynamical contributions are mixed by solving Eq.(\ref{Kapp}) for the scattering T-matrix. We do not attempt or even need to decompose artificially the derived T-matrix elements or cross sections into background and resonance components but only state, where meaningful, the net result. In all of the calculation, coupled channels effects are extremely important. The large contributions of higher order terms due to the repeated interactions in the summation of the scattering series can be seen quantitatively in various figures where the input Born terms are compared to the final result of the full T-matrix solution. The $P_{I1}$ partial wave cross section rises steeply, exceeding the contribution of other partial waves in the region from 1.8 to 1.95 GeV and is responsible for the steep rise of the total cross section at high energies.

\begin{figure}
  \begin{center}
{\includegraphics*[width=17cm]{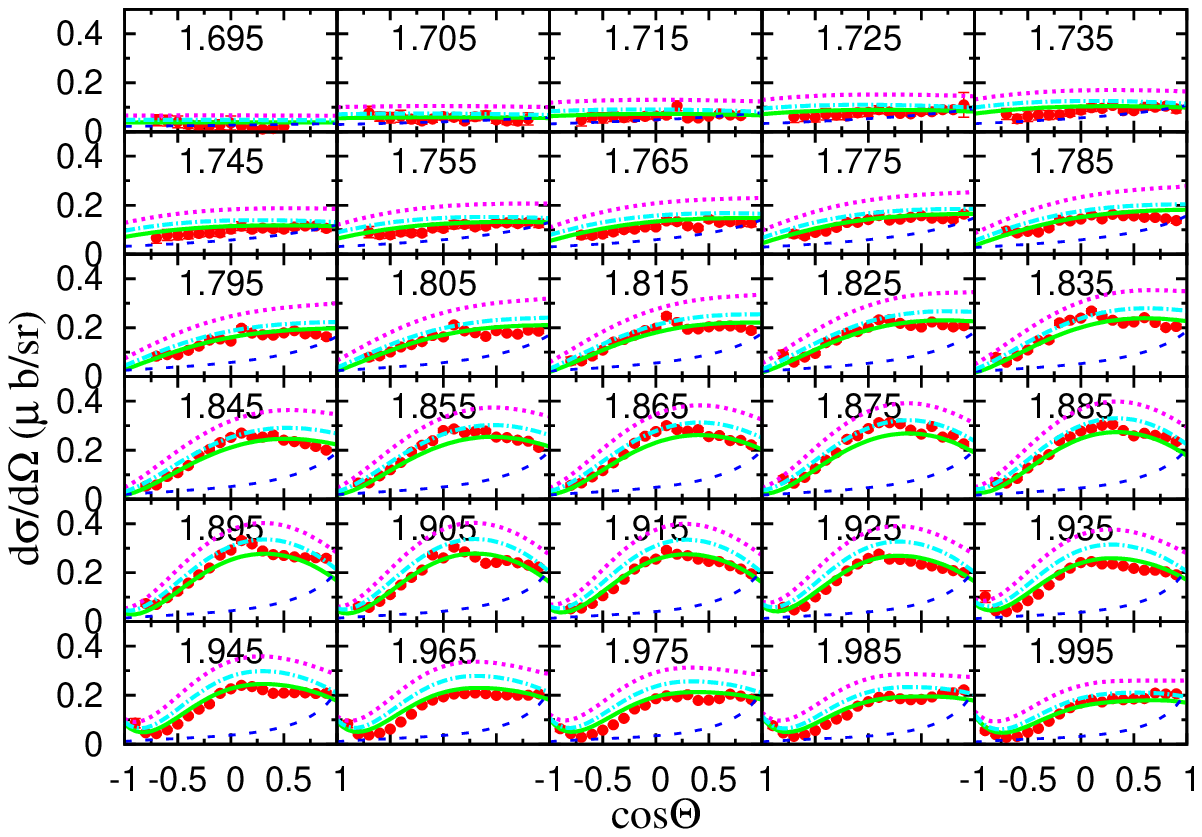}}
       \caption{
(Color online) The differential cross section of $\gamma p \to K^+ \Sigma^0$ reaction. The solid (green), dashed (blue), dotted (magenta) and dash-dotted (cyan) lines are the full model calculation, the contribution of the Born terms, the model calculation with the $S_{11}(1650)$ and $S_{31}(1620)$ turned off, respectively. The data are taken from CLAS~\cite{CLAS10Dey}. The numeric values label the center of mass energies in unit of GeV.
      \label{gs0dif}}
  \end{center}
\end{figure}

\begin{figure}
  \begin{center}
{\includegraphics*[width=17cm]{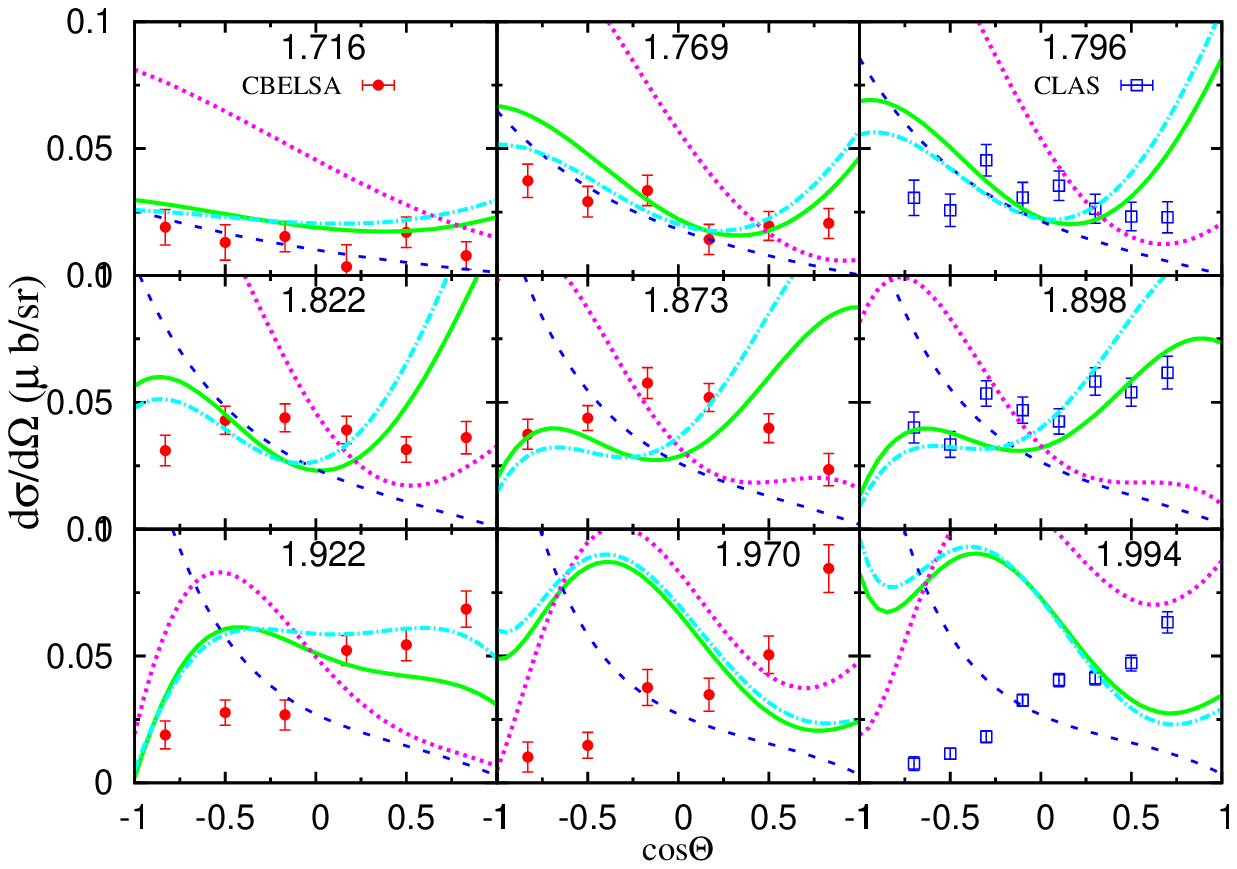}}
       \caption{
(Color online) The differential cross section of $\gamma p \to K^0 \Sigma^+$ reaction. The solid (green), dashed (blue), dotted (magenta) and dash-dotted (cyan) lines are the full model calculation, the contribution of the Born terms, the model calculation with the $S_{11}(1650)$ and $S_{31}(1620)$ turned off, respectively. The data are taken from CLAS~\cite{CLASthesis} and CBELSA~\cite{CBELSA08}. The numeric values label the center of mass energies in unit of GeV.
      \label{gspdif}}
  \end{center}
\end{figure}

It is known that the Born term is enhanced in the photo-induced reactions due to gauge invariance~\cite{Penner:2002b,Penner:thesis,Usov2005}. As a result, the contribution from the nucleon Born term to the $\gamma p \to K \Sigma$ reactions in our model is larger than the t-channel meson exchange. The Born term gives an important contribution to the $S_{I1}$ waves in the $K^+ \Sigma^0$ channel and to the $S_{I1}$, $P_{I1}$ and $P_{I3}$ waves in the $K^0 \Sigma^+$ channel, as shown in the right panels of Fig.~\ref{gstots}. The contribution from the Born term to the angular distributions are also shown in Fig.~\ref{gs0dif} and Fig.~\ref{gspdif}. It is a challenge to explain the recent $K \Sigma$ photoproduction data with the small total cross section of the $\gamma p \to K^0 \Sigma^+$ channel as compared to the $\gamma p \to K^+ \Sigma^0$ reaction. We find that the initial input of t-/u-channel background and s-channel resonance contributions interfere in the final T-matrix destructively, leading to the smaller total cross section for the $K^0 \Sigma^+$ channel, while it is constructive in $K^+ \Sigma^0$ channel. This difference is essential to suppress the total cross section of the $\gamma p \to K^0 \Sigma^+$ channel in our present model. Though the agreement to the CLAS and CBELSA data are still poor as depicted in Fig.~\ref{gstots}, it is clear that the data of the $K^0 \Sigma^+$ channel provides an additional constraint for the model parameters. As a result, the extracted coupling at the $N K \Sigma$ vertex changes its sign compared to the previous Giessen model, as shown in Tab.~\ref{borncoupling}.

Since the background contribution, rescattering and interference strongly influence the $\gamma p \to K \Sigma$ reactions, it is difficult to identify unambiguously individual resonance contributions only from the partial wave decomposition in Fig.~\ref{gstots}. In order to give an overall understanding of the production mechanism, we demonstrate the rating of the resonances in the $K \Sigma$ photoproduction in the last column of Tab.~\ref{resonance12} and Tab.~\ref{resonance32}, based on our present calculations. Here the rank is defined by the absolute variant value of the $\chi^2$ in the $\gamma p \to K \Sigma$ reactions after turning off the corresponding resonance. So the highest ranking three stars ($\Delta\chi^2 > 10$) represent the significant role of these resonances, and the two stars (10 $> \Delta\chi^2 > 5$) stand for the moderate contribution from the corresponding resonances. The one star states ($\Delta\chi^2 < 5$) play only a minor or no role in the $\gamma p \to K \Sigma$ reactions. As can be seen from the ranking, in the isospin $I = 3/2$ sector the $P_{31}(1750)$, $P_{33}(1600)$, $D_{33}(1700)$, $D_{35}(1930)$ and $F_{35}(1905)$ resonances contribute most noticeably to the $K \Sigma$ photoproduction. In the isospin $I = 1/2$ sector, the contributions of the $S_{11}(1650)$, $P_{11}(1440)$, $D_{13}(1520)$, $D_{15}(1675)$ and $F_{15}(1680)$ states are most significant.

The $S_{11}(1650)$ state plays an important role in the $S_{11}$ partial wave in both production channels at low energies. Close to threshold, the interference between $S_{11}(1650)$ and background develops a steep rise of the total cross section in the $K^0 \Sigma^+$ channel as seen in Fig.~\ref{gstots}. The kink structure around 1.72 GeV in the $S_{11}$ partial wave is due to the $\omega N$ production threshold. The $S_{31}(1620)$ state plays an important role in the $S_{31}$ partial wave which is however much smaller than $S_{11}$ channel. These two resonances, $S_{11}(1650)$ and $S_{31}(1620)$, have an obvious influence in the angular distributions of both channels, as shown in Fig.~\ref{gs0dif} and Fig.~\ref{gspdif}.

\begin{figure}
  \begin{center}
{\includegraphics*[width=17cm]{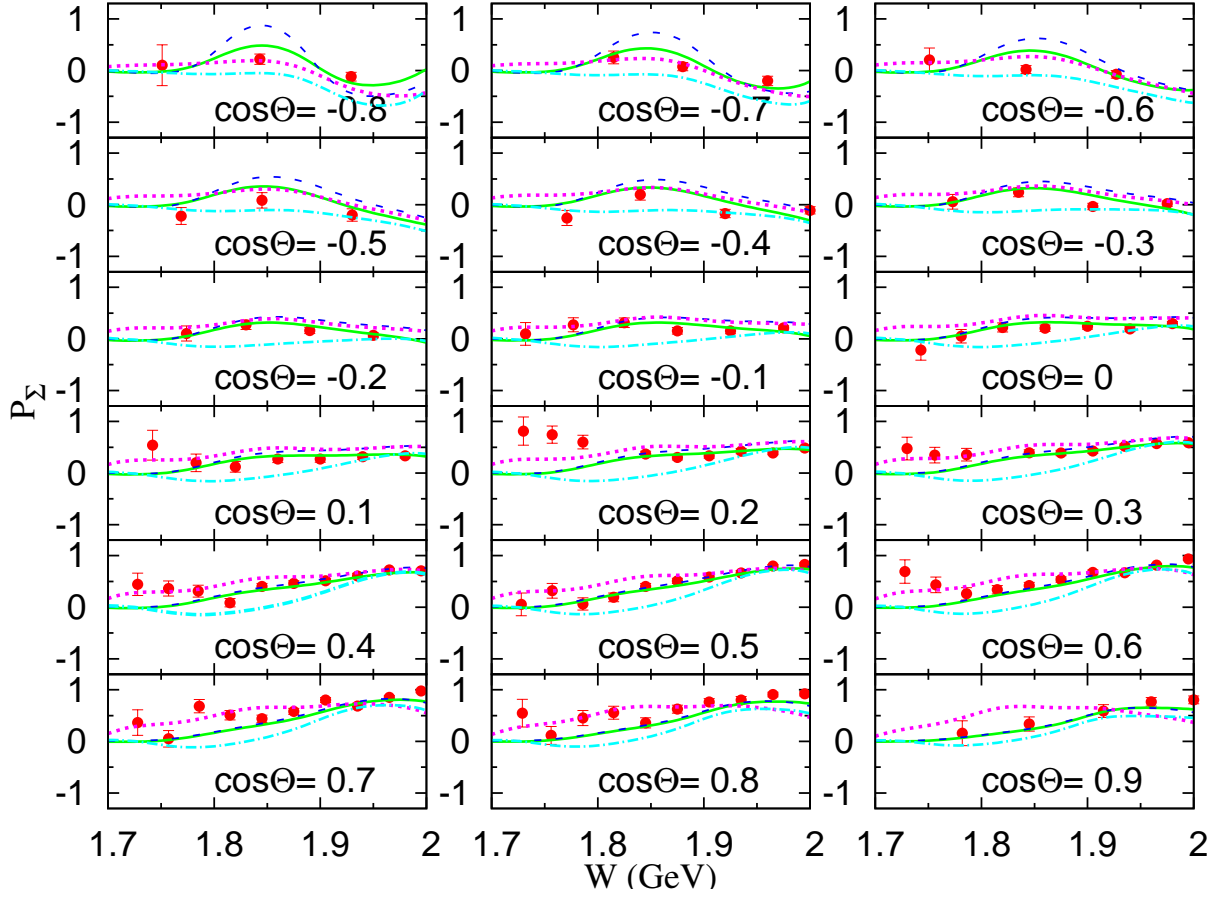}}
       \caption{
(Color online) The recoil polarization of $\gamma p \to K^+ \Sigma^0$ reaction. The solid (green), dotted (magenta), dash-dotted (cyan) and dashed (blue) lines are the full model calculation, the model calculation with the $P_{31}(1750)$, $D_{33}(1700)$ and $F_{15}(1680)$ turned off, respectively. The data are taken from CLAS~\cite{CLAS10Dey}.
      \label{gs0rec}}
  \end{center}
\end{figure}

\begin{figure}
  \begin{center}
{\includegraphics*[width=17cm]{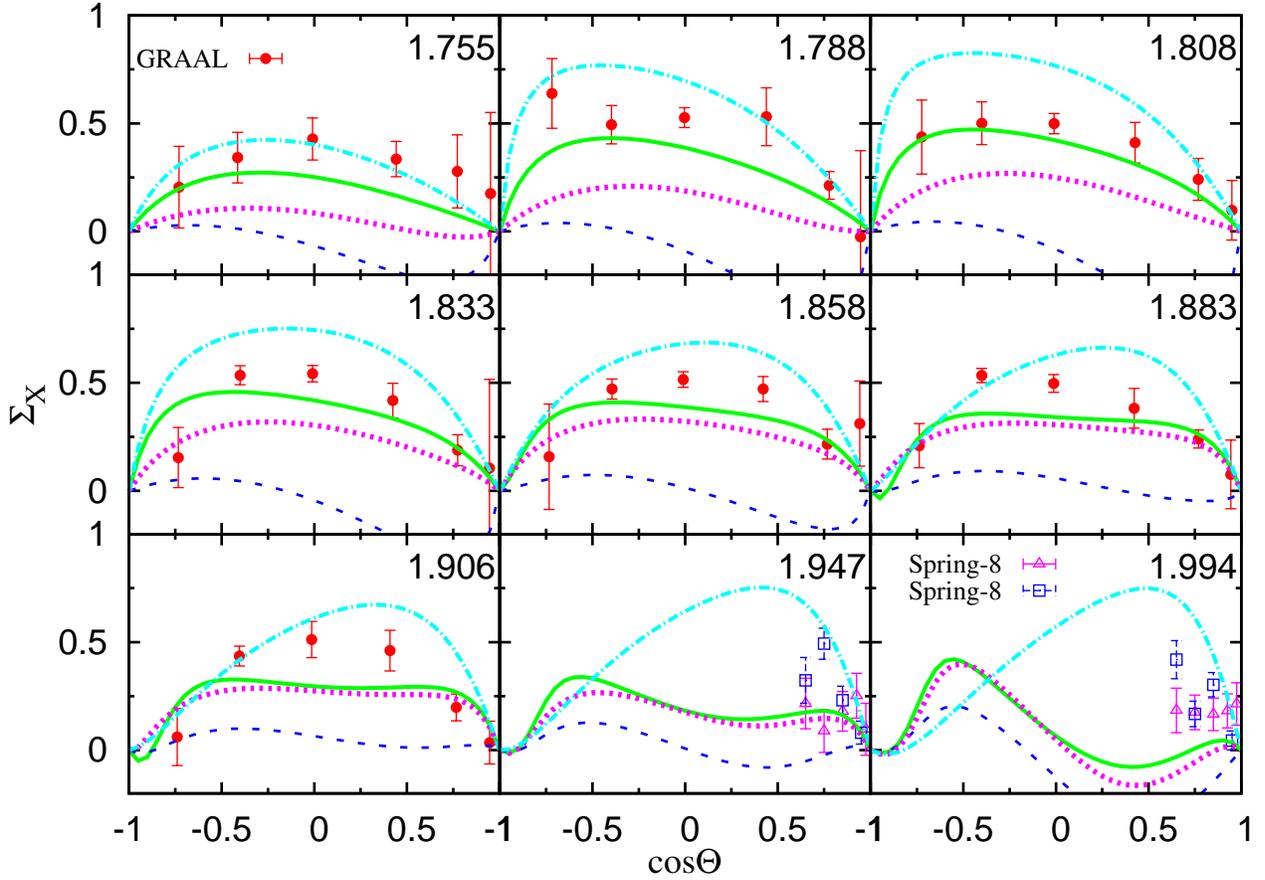}}
       \caption{
(Color online) The beam asymmetry of $\gamma p \to K^+ \Sigma^0$ reaction. The solid (green), dashed (blue), dotted (magenta) and dash-dotted (cyan) lines are the full model calculation, the model calculation with the $P_{31}(1750)$, $D_{33}(1700)$ and $D_{35}(1930)$ turned off, respectively. The data are taken from GRAAL~\cite{GRAAL07} and LEPS~\cite{LEPS03,LEPS06Kohri,LEPS06Sumihama}. The numeric values label the center of mass energies in unit of GeV.
      \label{gs0sig}}
  \end{center}
\end{figure}

\begin{figure}
  \begin{center}
{\includegraphics*[width=17cm]{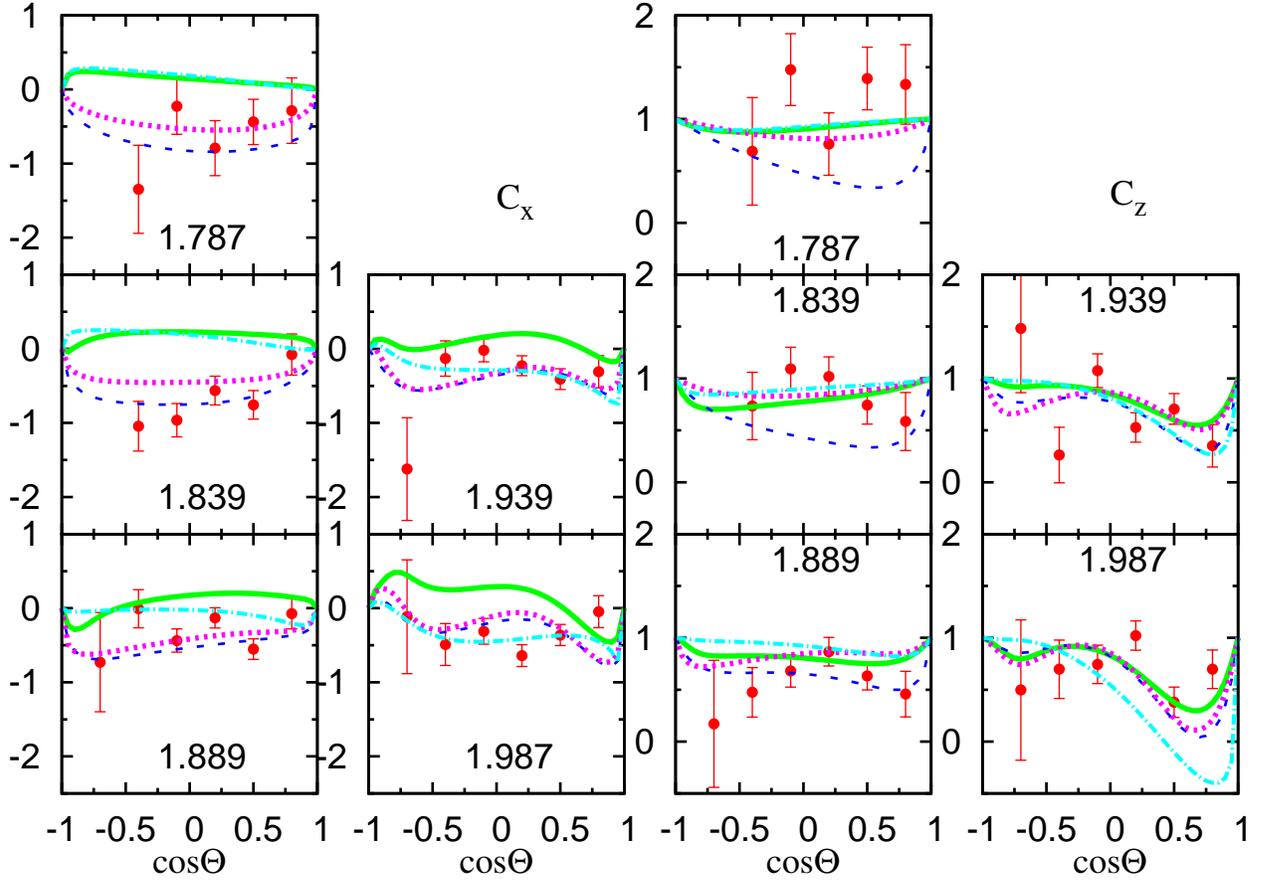}}
       \caption{
(Color online) The spin transfer coefficient $C_{x}$ and $C_{z}$ in $\gamma p \to K^+ \Sigma^0$ reaction. The solid (green), dashed (blue), dotted (magenta) and dash-dotted (cyan) lines are the full model calculation, the model calculation with the $P_{31}(1750)$, $D_{33}(1700)$ and $F_{35}(1905)$ turned off, respectively. The data are taken from CLAS~\cite{CLAS07CxCz}. The numeric values label the center of mass energies in unit of GeV.
      \label{gs0cxz}}
  \end{center}
\end{figure}

In the $K^+ \Sigma^0$ channel, the $P_{31}(1750)$ resonance is important for producing a broad shoulder in the total cross sections around 1.8 GeV, see lower panels of Fig.~\ref{gstots}. In the $K^0 \Sigma^+$ channel, the $P_{31}(1750)$ is important not only for the peak around 1.8 GeV but also for the steep rise at high energies. A closer inspection reveals that these contributions overestimate the total cross sections in both the channels below 1.85 GeV, as can be seen in Fig.~\ref{gstots}. In the $K^+ \Sigma^0$ channel, the $P_{31}(1750)$ contributes to the non-symmetric shape of the beam asymmetry up to 1.85 GeV and the bump around 1.85 GeV in the backward angles in the recoil polarization, see Fig.~\ref{gs0sig} and Fig.~\ref{gs0rec} respectively. It also reduces considerably the magnitude of the double polarization $C_{z}$ to the observed value in Fig.~\ref{gs0cxz}. In the recoil polarization of the $K^0 \Sigma^+$ channel, its effect can be seen but on a relatively low level.

The $D_{33}(1700)$ state is important for the shape of the beam asymmetry and recoil polarization in the $K^+ \Sigma^0$ channel. However, it shifts $C_{x}$ to the positive values which seems to be unfavored by data, as shown in the left panels of Fig.~\ref{gs0cxz}. Other resonances, i.e. $P_{33}(1600)$, $P_{33}(1920)$ and $S_{31}(1900)$, are important to bring the calculated $C_{x}$ back to the negative values as demanded by the CLAS data.

\begin{figure}
  \begin{center}
{\includegraphics*[width=17cm]{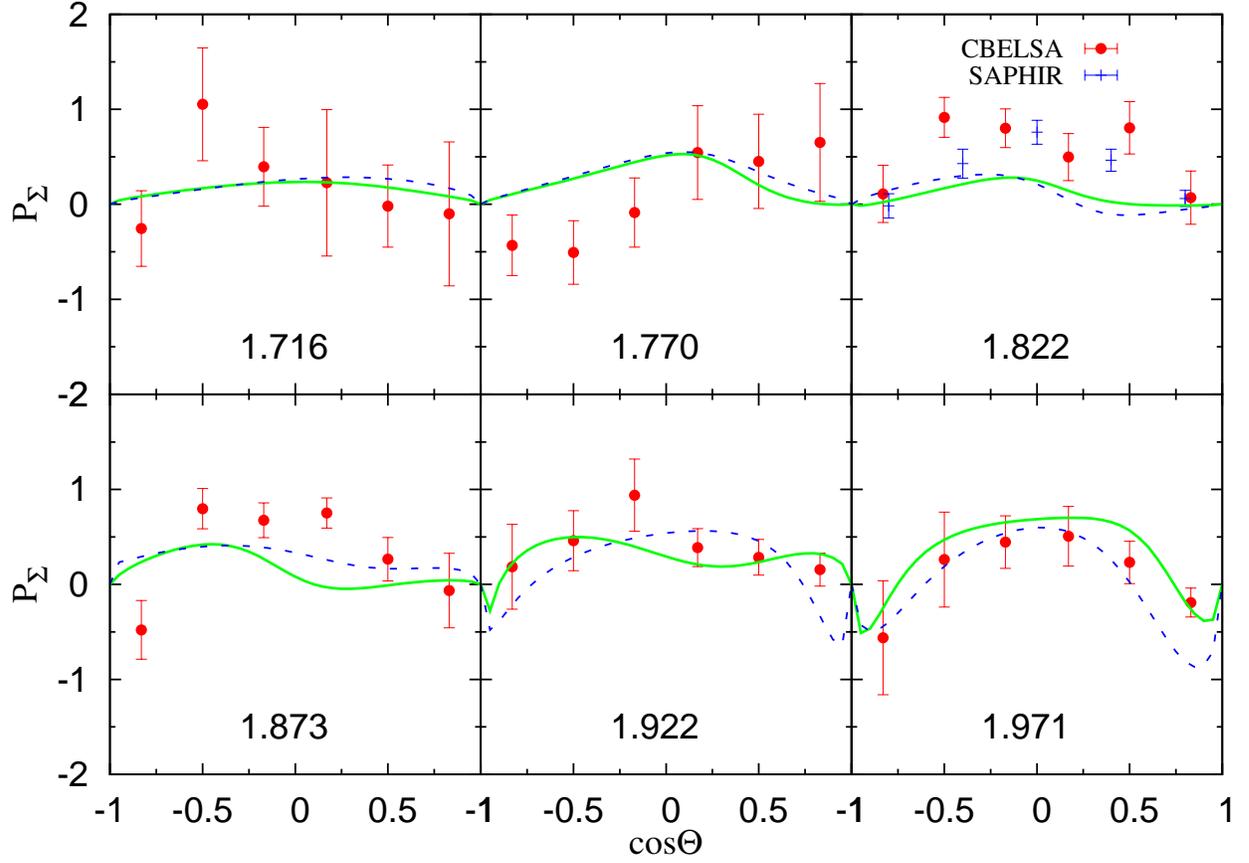}}
       \caption{
(Color online) The recoil polarization of $\gamma p \to K^0 \Sigma^+$ reaction. The solid (green) and dashed (blue) lines are the full model calculation and the model calculation with the $D_{15}(1675)$ turned off, respectively. The data are taken from CBELSA~\cite{CBELSA08} and SAPHIR~\cite{SAPH05}. The SAPHIR data is only for comparison but not used in the parameters fitting. The numeric values label the center of mass energies in unit of GeV.
      \label{gsprec}}
  \end{center}
\end{figure}

Similar to the $\pi N \to K \Sigma$ reactions, the $D_{15}(1675)$ and $D_{35}(1930)$ resonances are clearly reflected in all observables in the $K \Sigma$ photoproduction, as illustrated in Fig.~\ref{gsprec} and Fig.~\ref{gs0sig}. As shown in Fig.~\ref{gs0sig}, the $D_{35}(1930)$ is responsible for the backward peak in the beam asymmetry of the $K^+ \Sigma^0$ channel. The $F_{15}(1680)$ resonance is important for producing the backward structures in the angular distributions and polarization observables of both channels, as shown in Fig.~\ref{gs0rec}.

The angular distributions of $K^+ \Sigma^0$ channel in Fig.~\ref{gs0dif}, showing a remarkable enhancement at forward angles, is induced by many resonances. Although the shape originates mainly from the $P_{31}(1750)$, $D_{13}(1520)$, $D_{33}(1700)$ and $D_{35}(1930)$ resonances, the contribution from the $S_{11}(1650)$,  $P_{33}(1600)$ and $D_{15}(1675)$ are also very important for reproducing the total magnitude. At high energies, two higher lying resonances $P_{13}(1900)$ and $F_{35}(1905)$ participate. The shape of the angular distribution is a result of the interference of these resonances with the background amplitudes. These resonances are equally important also for the beam asymmetry and polarization observables of the $K^+ \Sigma^0$ channel. The beam asymmetry and recoil polarization in Fig.~\ref{gs0sig}, Fig.~\ref{gs0dif} and Fig.~\ref{gsprec} are nicely reproduced. The calculated spin transfer coefficient $C_{x}$ seems to be around zero but the CLAS data in the right panels of Fig.~\ref{gs0cxz} indicate an increased negative value so the fit quality of this observables needs improvement. As shown in the right panels of Fig.~\ref{gs0cxz}, the value of $C_{z}$ is close to one and is trivially explained in our model, because it is mainly determined by the Born term with additional small structures from resonances.

As pointed out above, the broad peak in the $K^+ \Sigma^0$ differential cross sections, see Fig.~\ref{gs0dif}, is produced by the interference of several resonances with background contributions. In the $K^0 \Sigma^+$ differential cross sections, where no obvious peak structure is observed in Fig.~\ref{gspdif}, the role of resonances and the background contributions still should be important. However, as seen in Fig.~\ref{gspdif}, the measured angular distributions are rather poorly described, becoming worse above 1.9 GeV. As shown in the upper-right panels of Fig.~\ref{gstots}, at high energies the $P_{I1}$ partial wave from background is contributing significantly to this channel. But the contribution of resonances is not enough to provide a destructive interference to compensate this $P_{I1}$-wave excess, which results in a poor description of both total and differential cross sections at high energies. In contrast, the recoil polarization data in Fig.~\ref{gsprec} seem to be structureless and flat within the experimental errors.

In the $\pi N \to K \Sigma$ reactions, it has been shown that the data of angular distributions and polarization observables provide plenty of information on the individual partial waves. This effect is seen clearly in the $K \Sigma$ photoproduction. Among the lowest rank states, the $P_{11}(1710)$, $F_{15}(2000)$ and $F_{35}(2000)$ have visible effects on the magnitude of the total and differential cross sections of the $K^0 \Sigma^+$ channel. Though the $F_{35}(1905)$ hardly affects the total cross sections, it influences considerably the double polarization observables of $K^+ \Sigma^0$ channel at high energies, for an illustration see Fig.~\ref{gs0cxz}. The one star state in our ranking, the $P_{33}(1232)$ and $P_{33}(1920)$ resonance, are seen in the angular distributions and polarization observables of both channels. The $D_{13}(1950)$ state seems to be of minor importance and it is the only resonance that is hardly seen in all observables.

\subsection{Discussion}

From the present analysis of the new $K \Sigma$ photoproduction data we have obtained stringent constraints on the resonance couplings in our Lagrangian. The values of the coupling constants for the strangeness-carrying vertices derived here are compared to the previously obtained results \cite{Penner:2002a} in Tab.~\ref{borncoupling}. The least change is found in the $NK_0^{*}\Sigma$ couplings which in magnitude are increased by about 20\%. The $NK\Sigma$ coupling constants are altered more drastically: besides a change of sign their magnitude is increased by a factor of more than 2. This sign change is important because it leads the small total cross sections of the $\gamma p \to K^0 \Sigma^+$ channel, as pointed out in Sec~\ref{KSigmaphoto}. There is always an ambiguity in the sign of different couplings constants; for example the definition of $N K \Sigma$ coupling in Ref.~\cite{Shyam10} differs from ours in sign. However, in SU(3) symmetry the relative sign of $N K \Sigma$ and $N K \Lambda$ is negative~\cite{Saghai1995}. The $N K \Lambda$ coupling extracted earlier~\cite{Shklyar05lam} is of the same sign as the present $N K \Sigma$ coupling, indicating that our annalysis does not follow the SU(3) relations. Similar result was found also in the Usov\&Scholten approach~\cite{Usov2005,Shyam10} as well as in other models~\cite{Saghai1995}. It should be noted that our negative $N K \Lambda$ coupling is extracted from the data published before 2007, so it is interesting to check this sign problem after including the enlarged polarization data on $K \Lambda$ photoproduction into the analysis. Also, the $NK^*\Sigma$ and $NK_1\Sigma$ couplings constants are significantly modified: we obtain much smaller values than in \cite{Penner:2002a}. As a result, the contributions of the t-channel $K^*$ and $K_1$ meson exchange are decreased, while the strength of the $K_0^{*}$ part is slightly increased.

The couplings of some resonances to the $K \Sigma$ channel are very small and even approaching zero, but this does not mean that they have no influence in the $K \Sigma$ production. They still can contribute through coupled-channel and interference effects. That, in fact, is an important reason why the model demands many resonances but only few of them, i.e. the $D_{13}(1520)$, $D_{15}(1675)$, $D_{35}(1930)$ and $F_{15}(1680)$ states, have large couplings to the $K \Sigma$ final channel. As seen in Tab.~\ref{resonance12}, the coupling constant of $P_{11}(1710)$ to the $K \Sigma$ channel is much smaller than in the previous Giessen analysis, so its contribution is suppressed. In the isospin $I = 3/2$ sector, the sign of the coupling constants of the $S_{31}(1620)$ and $P_{33}(1600)$ resonances to the $K \Sigma$ channel is opposite to those of the previous Giessen model as can be seen in Tab.~\ref{resonance32}. Also the electromagnetic helicity amplitudes of $P_{31}(1750)$ state is much smaller than in the previous investigations (see Tab.~\ref{resonance32}). Hence, the contribution of this state to the $K \Sigma$ photoproduction is decreased compared to our previous analysis. This effect is more pronounced at high energies.

In the Bonn-Gatchina partial wave analysis, the $S_{11}(1535)$, $S_{11}(1650)$, $P_{13}(1720)$ and $P_{11}(1840)$ states give the main contributions to hyperon photoproduction~\cite{BoCh07}. Especially the $C_{x}$ and $C_{z}$ observables in the $K^+ \Sigma^0$ channel require an additional $P_{13}(1900)$ resonance, as stressed in their analysis~\cite{BoCh08}. In our model, however, we include this resonance into the formulation from the very beginning~\cite{Penner:2002a,Penner:2002b} and in this paper we confirm its importance in the $K \Sigma$ photoproduction, especially at high energies. In a covariant isobar model~\cite{Mart2012} this resonance is also found to be important in producing the cross section peak around 1.9 GeV in $\gamma p \to K^+ \Lambda$. However, in the Giessen model this peak is caused by the interference effect of $P_{13}$ resonance and background terms~\cite{Shklyar05lam}. In the Bonn-Gatchina model, a high lying $P_{11}$ state with a mass of 1840 MeV is found to be important for the $K \Sigma$ photoproduction, and a third $S_{11}$ state with the mass around 1900 MeV is also needed in the global fit though only weakly contributing to $K \Sigma$ photoproduction. In order to check that conclusion we had added separately and arbitrarily a $S_{11}$, a $S_{31}$, a $P_{11}$ and a $P_{31}$ resonance with BW mass varying from 1700 Mev to 2000 MeV to our model, but without finding any evidence for such a contribution.

A puzzling result is the difference in the description of the $K^0\Sigma^+$ and the $K^+\Sigma^0$ photoproduction data. At practically all energies the $K^+\Sigma^0$ data are well described, including angular distributions of cross sections and polarization observables. In the complementary channel $K^0\Sigma^+$ we achieve, however, the measured observables are considerably less accurately reproduced, e.g. Fig.\ref{gspdif}. The differences are showing up most clearly in the differential observables, indicating remaining uncertainties in phase relations, obviously affecting the resulting interference pattern. Taking the valence quark configuration as a guideline the two channels differ only by the final distribution of $u$ and $d$ quarks among the two hadrons: the reaction $\gamma p\to K^0\Sigma^+$ corresponds to $K^0(d\overline{s})+\Sigma^+(uus)$ while the reaction $\gamma p\to K^+\Sigma^0$ is leading to $K^+(u\overline{s})+\Sigma^0(uds)$. Assuming charge symmetry at the quark (and hadron) level, the two channel configurations should behave perfectly the same, except for particular threshold contributions or resonances coupling differently to the two exit channels.

In our model the driving force for the population of the $K^0 \Sigma^+$ channel is the nucleon-photon Born term while in  the Bonn-Gatchina model, the main contribution  is the t-channel $K$ meson exchange~\cite{BoCh07}.  It should be mentioned that their agreement with $K^0 \Sigma^+$ data is worse than that for the $K^+ \Sigma^0$ channel, and their fit of $C_{x}$ leads to a $\chi^2$ value of little less than 3.0, larger than that of $C_{z}$~\cite{BoCh11}. These findings are in line with the results of our model so it seems that the $\gamma p\to K^0\Sigma^+$ reaction really needs further study in the future.

The recent analysis performed by Shyam et al.~\cite{Shyam10} within the Usov\&Scholten model~\cite{Usov2005} obtains a good description of the SAPHIR data but not the CLAS data. In that analysis the $\gamma p \to K^+ \Sigma^0$ reaction is dominated by the background and the $P_{33}(1600)$ resonance predicting a much simpler production mechanism. It should be pointed out that in Ref.~\cite{Shyam10} the $\gamma p \to K^0 \Sigma^+$ channel, which poses strong constraints to the model parameters in our extended approach, was not included into the analysis.

\begin{table}[t]
  \begin{center}
    \begin{tabular}
      {ccccccc }
      \hline
      $N^*$
& BW mass$^b$& $\Gamma_{tot}^{\quad b}$      & $R_{K\Sigma}$     & $R_{K\Sigma}^{\quad p}$ &  Rank \\
      \hline
   $S_{11}$(1535)
& 1526 & 135 &  $-0.64^a$ & $0.83^a$ & ** \\
   $S_{11}$(1650)
& 1664 & 119 &  $-0.81^a$ & $-0.59^a$ & *** \\
 \hline
  $P_{11}$(1440)
& 1517 & 608 &  $-0.45^a$ &  $0.53^a$ & *** \\
  $P_{11}$(1710)
& 1723 & 403 &   $0.1(-)$ &  $12.6(-)$ & * \\
 \hline
   $P_{13}$(1720)
& 1700 & 154 &  $ 0.0(-)$ &  $ 0.0(-)$ & ** \\
   $P_{13}$(1900)
& 1998 & 401 &  $ 0.4(+)$ & $ 2.0(-)$ & ** \\
 \hline
   $D_{13}$(1520)
& 1505 & 103 &  $1.35^a$ & $1.13^a$ & *** \\
   $D_{13}$(1950)
& 1935 & 858 & $ 0.0(-)$ & $ 0.3(+)$ & * \\
 \hline
   $D_{15}$(1675)
& 1665 & 147 & $ 5.4(+)$ &  --- & *** \\
 \hline
   $F_{15}$(1680)
& 1676 & 112 &  $65.9(-)$ &  ---  & ** \\
   $F_{15}$(2000)
& 1946 & 197 &  $ 0.0(+)$ &  --- & ** \\
      \hline
    \end{tabular}
  \end{center}
  \caption{Branching decay ratios $R_{K\Sigma} =\Gamma_{K\Sigma}/\Gamma_{tot}$ of $I = 1/2$ resonances into the  $K\Sigma$ final state extracted in the present calculation. In brackets, the sign of corresponding coupling constant is shown (all $\pi N$ couplings are chosen to be positive, see \cite{Shklyar04J}). The BW masses and total width are given in MeV and the decay ratios $R_{K\Sigma}$ in percent.\\
    $^a$: the coupling is given since the resonance BW mass is below the threshold.\\
    $^b$: fixed in the previous calculations \cite{Shklyar04J,Shklyar05ome,Shklyar07eta}.
    $^p$: the C-p-$\gamma +$ results from a previous Giessen model analysis \cite{Penner:2002a}.
    \label{resonance12}}
\end{table}

\begin{table}[t]
  \begin{center}
 \begin{tabular}{ccccccccc}
 \hline
 $\Delta^*$  & BW mass & $\Gamma_{tot}$ &
 $R_{\pi N}$ & $R_{2\pi N}$  & $R_{K\Sigma}$ &
 $A_{\frac{1}{2}}$  &   $A_{\frac{3}{2}}$ & Rank \\
 \hline
  $S_{31}$(1620)
& 1598 & 209 &  26.8 & $ 73.2(+)$ &  $-0.35^a$ &  -58 & --- & ** \\
& 1611 & 196 &  34.3 & $ 65.7(-)$ &  $ 0.14^a$ &  -50 & --- \\
 \hline
   $P_{31}$(1750)
& 1773 & 651 &   1.6 & $ 97.5(+)$ &  $ 0.9(+)$ &    1 & --- & ***\\
& 1712 & 660 &   0.8 & $ 99.1(+)$ &  $ 0.1(+)$ &   53 & --- \\
 \hline
   $P_{33}$(1232)
& 1227 & 110 & 100.0 & $  0.0(-)$ &  $ 0.00^a$ & -128 & -253 & ** \\
& 1228 & 106 & 100.0 & $  0.21(-)^b$ &$ 0.00^a$& -128 & -247 \\
   $P_{33}$(1600)
& 1694 & 515 &  14.9 & $ 85.1(+)$ &  $-0.10^a$ &  -10 &  -17 & *** \\
& 1667 & 407 &  13.3 & $ 86.7(+)$ &  $ 0.03^a$ &   0  &  -24 \\
   $P_{33}$(1920)
& 2069 & 767 &   4.1 & $ 95.1(-)$ & $ 0.7(-)$  &   21 &   25 & **\\
& 2057 & 494 &  15.9 & $ 81.6(-)$ & $ 2.4(-)$  &   -7 &   -1 \\
 \hline
   $D_{33}$(1700)
& 1673 & 766 &  15.0 & $ 85.0(+)$ & $ 0.17^a$  &  97  &  147 & ***\\
& 1678 & 591 &  13.9 & $ 86.1(+)$ & $ 0.75^a$  &  96  &  154 \\
 \hline
 $D_{35}$(1930)
& 2001 & 440 &   7.2 & $ 78.8(+)$ & $14.1(+)$   &  -66 &   1 & ***\\
 \hline
   $F_{35}$(1905)
& 1842 & 619 &   6.5 & $ 93.4(-)$ & $ 0.003(-)^b$  &   54 & -127 & ***\\
   $F_{35}$(2000)
& 2160 & 313 &   1.5 & $ 98.5(-)$ & $ 0.89(-)^b$  &   18 &  -23 & *\\

 \hline
   \end{tabular}
  \end{center}
  \caption{The first and second line are the properties of $I = 3/2$ resonances extracted in the present calculations and the C-p-$\gamma +$ results from a previous Giessen model analysis \cite{Penner:2002a}, respectively. The BW masses and total width are given in MeV and the decay ratios $R_{ab} =\Gamma_{ab}/\Gamma_{tot}$ in percent. The electromagnetic helicity amplitudes $A_{\frac{1}{2}}$ and $A_{\frac{3}{2}}$ are in the unit of $10^{-3}$ GeV$^{-1/2}$\\
    $^a$: the coupling is given since the resonance BW mass is below the threshold.
    $^b$: decay ratio in 0.01\%.
    \label{resonance32}}
\end{table}

\begin{table}[t]
  \begin{center}
  \begin{tabular}{c|c|c|c|c|c}
 \hline
  $ g_{NK\Sigma}$ & $g_{NK_0^{*}\Sigma}$ & $g_{NK^*\Sigma}$ & $\kappa_{NK^*\Sigma}$ & $g_{NK_1\Sigma}$ &  $\kappa_{NK_1\Sigma}$ \\
 \hline
  -5.41     &    -32.94 &   0.83  &    -1.71  &   3.67 & -2.58 \\
   2.48     &    -26.15 &   4.33  &    -0.86  &  22.80 &  2.40 \\
 \hline
 \end{tabular}
  \end{center}
  \caption{Born couplings in the present calculations (first line) compared to previous Giessen model (second line)~\cite{Penner:2002a}.
    \label{borncoupling}}
\end{table}

\section{Summary} \label{summary}

In the present paper, we perform a coupled-channel analysis which uses effective Lagrangian and respects unitary and gauge invariance to the $(\pi,\gamma) N \to K\Sigma$ reactions up to the center of mass energy of 2.0 GeV. The available data of pion- and photon-induced reactions are simultaneously analyzed to investigate the reaction mechanism. The meson-baryon coupling constants and resonance couplings to the $K\Sigma$ state are extracted. Several resonances contribute to the process. The coherent sum of resonances and background contributions is essential to describe the recent photoproduction data obtained by the CLAS, CBELSA, LEPS, and GRAAL groups. It has been shown that the $K\Sigma$ production mechanism is much more complicated than that concluded from the previous Giessen model studies after taking into account these new data. Overall, our results agree well with the data. However, there are puzzling exceptions, namely the double polarization spin transfer coefficients in the $\gamma p \to K^+ \Sigma^0$ and the differential cross sections of the $\gamma p \to K^0 \Sigma^+$ which are awaiting further investigation. In our planned model improvements and reformulations, one of the crucial directions is to treat the $2\pi N$ channels as the real $\rho N$, $\pi \Delta$ and $\sigma N$ states, which is under progress and will be the topic of separate publications. On the other hand, extrapolations into the complex plane and extracting the poles and residues of the full amplitudes are of fundamental interest and should be considered as a major direction in the future. In summary, we find that $K \Sigma$ production is a good probe to explore the isospin $I = 3/2$ resonances. Our results are shedding light on the search for missing resonances and the $K\Sigma$ production mechanism in other reactions, for example, the long standing controversy in the close-to-threshold behavior of $pp \to n K^+ \Sigma^+$ reaction~\cite{caoppnKSigma}.

\begin{acknowledgments}

We thank Dr. O. Jahn and Prof. U. Thoma for providing the CB-ELSA data. The effort from Ms. J. Yang in preparing the input data files is gratefully acknowledged. This work was supported by the
Deutsche Forschungsgemeinschaft (CRC16, grant B7 and grant Le439/7) and in part by I3HP SPHERE.

\end{acknowledgments}

\end{document}